\documentclass[12pt]{article}
\pdfoutput=1
\usepackage{subfigure}
\usepackage{amssymb,amsmath}
\usepackage{graphicx}
\usepackage{color}
\usepackage{cancel}
\usepackage[colorlinks=true
,urlcolor=blue
,citecolor=blue
,linkcolor=blue
,pagecolor=blue
,linktocpage=true
,pdfproducer=medialab
]{hyperref}
\usepackage[a4paper,width=15.2cm]{geometry}

\usepackage{ulem}

\makeatletter \renewcommand{\@dotsep}{10000} \makeatother
\def\be{\begin{equation}}
\def\ee{\end{equation}}
\def\bea{\begin{eqnarray}}
\def\eea{\end{eqnarray}}
\def\bi{\begin{itemize}}
\def\ei{\end{itemize}}

\newcommand\prd[3]{{\it Phys.\ Rev.\ }{\bf D #1} (#2) #3}
\newcommand\prl[3]{{\it Phys.\ Rev.\ Lett.\ }{\bf #1} (#2) #3}

\newcommand{\beq}{\begin{equation}}
\newcommand{\eeq}{\end{equation}}

\begin{document}
%Remove date before submitting to arXi

\date{\today}

\begin{center}
{\Large\bf 
Stop on Top: \\ \vspace{0.3cm}
SUSY Parameter Regions, Fine-Tuning Constraints 
} \vspace{1cm}
\end{center}

\begin{center}

{\Large
Durmu{\c s} Ali Demir$^{a}$\footnote{
Email: demir@physics.iztech.edu.tr}
and
Cem Salih \"{Un}$^{b}$\footnote{
Email: cemsalihun@uludag.edu.tr}
}

\vspace{0.75cm}

{\it $^a$
Department of Physics, {\.I}zmir Institute of Technology, IZTECH, TR35430, {\. I}zmir, Turkey
}\\
{\it  $^b$
Department of Physics, Uluda{\~g} University, TR16059, Bursa, Turkey}

\vspace{1.5cm}
\section*{Abstract}
\end{center}

We analyze minimal supersymmetric models in order to determine in what 
parameter regions with what amount of fine-tuning they are capable of 
accomodating the LHC-allowed top-stop degeneracy window. The stops must 
be light enough to enable Higgs naturalness yet heavy enough to induce 
a 125 GeV Higgs boson mass. These two constraints imply a large mass splitting.  

By an elaborate scan of the parameter space, we show that stop-on-top 
scenario requires at least $\Delta_{CMSSM} \simeq {\mathcal{O}}(10^{4})$ 
fine-tuning in the CMSSM. By relaxing the CMSSM parameter space with 
nonuniversal Higgs masses, we find that $\Delta_{NUHM1} 
\simeq {\mathcal{O}}(10^{4})$. The CMSSM with gravitino LSP works 
slightly better than the NUHM1 model. Compared to all these, 
the CMSSM with $\mu<0$ and nonuniversal gauginos yield a much 
smaller fine-tuning $\Delta_{\mu,g} \simeq {\mathcal{O}}(100)$. Our
results show that gaugino sector can pave the road towards a
more natural stop-on-top scenario. 

\newpage

%%%%%%%%%%%%%%%%%%%%%%%%%%%%%%%%%%%%%%%%%%%%%%%%%%%%%%%%%%%%
\renewcommand{\thefootnote}{\arabic{footnote}}
\setcounter{footnote}{0}

%%%%%%%%%%%%%%%%%%%%%%%%%%%%%%%%%%%%%%%%%%%%%%%%%%%%%%%%%%%%%

%\baselineskip 36pt
% Main body
%%%%%%%%%%%%%%%%%%%%%%%%%%
%\baselineskip 18pt
%%%%%%%%%%%%%%%%%%%%%%%%%%

%%%%%%%%%%%%%%%%%%%%%%%%%%%%
\section{Introduction}
\label{sec:intro}
%%%%%%%%%%%%%%%%%%%%%%%%%%%%

A new bosonic resonance of mass about 125 GeV has recently been discovered by ATLAS \cite{ATLAS} and CMS \cite{CMS} experiments at CERN. Even though the current analyses show that this new resonance exhibits properties very similar to the Standard Model (SM) Higgs boson, there is no doubt that the SM is not the final description of nature due to its drawbacks such as gauge hierarchy problem \cite{Gildener:1976ai} and absolute stability of of the SM Higgs potential \cite{hinstability}. The 125 GeV Higgs boson can give hints of new physics beyond the SM, and here supersymmetry (SUSY) stands out as one of the forefront candidates. The minimal supersymmetric extension of the SM (MSSM) can resolve the gauge hierarchy problem by invoking superpartners of the SM fields. Also, the lightest CP-even Higgs boson, one of the five physical Higgs states of the MSSM, exhibits very similar properties as the SM's Higgs boson in the decoupling limit \cite{Haber:1993pv}. In addition, imposing R-parity conservation the MSSM gives a highly plausible candidate for the dark matter in the lightest supersymmetric particle (LSP). 

Nevertheless, the LHC, as hadron smasher experiments, has brought very severe bounds on the color sector. Since there is no significant deviation in the Higgs production and decay properties with respect to the SM, and no significant signal of the supersymmetric particles, results from ATLAS and CMS have lifted the lower bounds on masses of gluinos and squarks in the first two generations up to $\sim 1.7$  TeV \cite{TheATLAScollaboration:2013fha}. When squarks of the first two generations are heavy and decoupled, the bound on gluino mass is lowered depending on its decay channels. For instance, through gluino-mediated pair production of the third generation squarks, gluinos of mass up to $\sim 1 - 1.3$ TeV are excluded depending on the final states \cite{TheATLAScollaboration:2013fha,GOUSKOS:2013xsa}.  Even though ATLAS and CMS results are presently not much severe for the third generation squarks, stop masses in the range of $100-750$ GeV have been excluded for massless LSP \cite{LARI:2014jia} (bounds on the stop mass are relaxed for massive LSP). The stop decay channels  $\tilde{t}\rightarrow t\tilde{\chi}_{1}^{0}$ and $\tilde{t}\rightarrow b\tilde{\chi}_{1}^{\pm}$ exclude stop masses up to 650 GeV \cite{Outschoorn:2013pma}. Moreover, ATLAS collaboration has recently looked for NLSP stop via its decay mode to charm quark and LSP, and the results have excluded stop masses up to 230 GeV \cite{TheATLAScollaboration:2013aia}.  This channel also puts a lower bound on gluino mass as $m_{\tilde{g}} \gtrsim 1.1 $ TeV \cite{COTE:2014yha}.

These exclusion model-specific limits express non-observation of light stops and gluinos at the LHC. The exclusion is not entire. The reason is that a small region with $m_{\tilde{t}_{1}} \lesssim 200$ GeV in $m_{\tilde{\chi}_{1}^{0}}-m_{\tilde{t}_{1}}$ plane has not been excluded yet. Indeed, discrimination of $t\bar{t}$ and $\tilde{t}\tilde{t}^{*}$ events with identical final states in this region is challenging \cite{Buckley:2014fqa} and $\tilde{t}\tilde{t}^{*}$ cross section stays in the error bar in calculation of top pair production \cite{Li:2013uma} which is measured to be \cite{Collaboration:2013bma}
\begin{equation}
\sigma_{t\bar{t}}^{\sqrt{s}=8 ~{\rm TeV}} = 241\pm 2~({\rm stat.}) \pm 31 ~({\rm syst.})\pm 9~({\rm lumi.})\, {\rm pb} 
\end{equation}

As is well-known for a long time, the Higgs boson mass is bounded from above by $M_{Z}$ at tree-level in the MSSM, and hence, one definitely needs to utilize radiative corrections in order to have 125 GeV Higgs boson mass. Since the Yukawa couplings for the first two families are negligible, the third family stands out as the dominant source inducing such sizeable quantum contributions to Higgs boson mass. The sbottom contribution is proportional to sbottom mixing parameter and $\mu \tan\beta$. However, strong bounds from vacuum stability on the $\mu\tan\beta$ term allows only a minor contribution from the sbottom sector \cite{Carena:2012mw}. Thus, the 125 GeV Higgs boson largely constrains the stop sector. In the case of $m_{\tilde{t}_{L}} \simeq m_{\tilde{t}_{R}}$, the left- and right-handed stop masses are excluded up to $\sim 800$ GeV. If one sets a hierarchy between the two stops ($m_{\tilde{t}_{L}} \ll m_{\tilde{t}_{R}}$ or vice versa), then it becomes possible to have a light stop, while the other must weigh above $\sim 1$ TeV for moderate $\tan\beta$ \cite{Carena:2011aa}. This hierarchy necessitates a large stop mixing.  Since MSSM has many scalar fields, one may be concerned about color and/or charge breaking (CCB) minima that could occur in case of large mixings at which some scalar fields may develop non-zero vacuum expectation values (VEVs). Even though large $m_{0}$ can ensure absence of such minima \cite{Ellwanger:1999bv}, specifically the sfermions of the third family need a careful treatment, since their VEVs may cause tunneling into a deeper CCB minimum \cite{Camargo-Molina:2013sta}. Among the MSSM scalars, stop has a special importance, since its non-zero VEV breaks $SU(3)_{c}$ and $U(1)_{{\rm em}}$ both. We thus check the vacuum stability for the CMSSM benchmark points listed in the text. Apart from semi-analytic estimates, the public code Vevacious \cite{Camargo-Molina:2013qva} returns stable vacua for certain parameter regions accommodating light stops. The stability of the CMSSM points encourages us to conclude stability of more relaxed, less constrained SUSY models.

Returning to the large mixing, a 125 GeV Higgs boson in the MSSM necessitates large splitting between the two stops, and this obviously contradicts with the naturalness domain $m_{\tilde{t}_{1}},~m_{\tilde{t}_{2}},~m_{\tilde{b}_{1}} \lesssim 500$ GeV \cite{Kitano:2005wc} unless some extensions of MSSM are considered \cite{Wymant:2012zp}. From the naturalness point of view, one thus concludes that light stop regions in the MSSM need significant fine-tuning to yield the electroweak scale ($M_{{\rm EW}} \sim 100$ GeV) correctly. In order to analyze the amount of fine-tuning in allowed parameter regions, one can specifically focus on the Z-boson mass ($M_{{\rm Z}} = 91.2$ GeV) 
\begin{equation}
\frac{1}{2}M^2_{{\rm Z}} = -\mu^{2}+\frac{(m^{2}_{H_{d}}+\Sigma_{d}^{d})-(m^{2}_{H_{u}}+\Sigma_{u}^{u})\tan^{2}\beta}{\tan^{2}\beta -1}
\label{ew-ft}
\end{equation}
which follows from the minimization of the MSSM Higgs potential \cite{Martin:1997ns} such that $\mu$ is the Higgsino Dirac mass, $\tan\beta = \langle H_u^0 \rangle/\langle H_d^0 \rangle$, 
$\Sigma^{u}_{u}$, $\Sigma^{d}_{d}$ are radiative effects from the Higgs potential, and $m^2_{H_{u,d}}$ are the soft-masses of the Higgs doublets $H_{u,d}$ which give mass to u-type and 
d-type fermions. 

For quantifying the amount of fine-tuning associated to $M_{{\rm EW}}$, we utilize the measure defined in the recent work \cite{Baer:2012mv}. Namely, we introduce the electroweak 
fine-tuning 
\begin{equation}
\Delta_{EW} \equiv {\rm Max}(C_{i})/(M_{{\rm Z}}^{2}/2)
\label{DeltaEW}
\end{equation}
where
\begin{equation}
C_{i}\equiv \left\lbrace 
\begin{array}{c}
\hspace{-1.2cm}C_{H_{d}}=\mid m^{2}_{H_{d}}/(\tan^{2}\beta -1) \mid \\ \\
C_{H_{u}}=\mid m^{2}_{H_{u}}\tan^{2}\beta/(\tan^{2}\beta -1) \mid \\ \\
\hspace{-3.8cm}C_{\mu}=\mid -\mu^{2}\mid
\end{array}
\right.
\end{equation}
follow from Higgs potential whose parameters are evaluated at the electroweak scale. 
%The fine-tuning can be continued to high-energy scales by running (\ref{ew-fit}) appropriately. Namely, one has at high scales
%\begin{equation*}
%\frac{1}{2}M_{{\rm Z}} =\frac{(m^{2}_{H_{d}}(M_{{\rm HS}})+\delta m^{2}_{H_{d}}+\Sigma_{d}^{d})-(m^{2}_{H_{u}}(M_{{\rm HS}})+
%\delta m^{2}_{H_{u}}+\Sigma_{u}^{2})\tan^{2}\beta}{\tan^{2}\beta -1} 
%\end{equation*}
%\begin{equation}\hspace{-6.0cm}
%- (\mu^{2}(M_{{\rm HS}})+\delta \mu^{2})\,. 
%\label{hs-ft}
%\end{equation}
%and similarly

The fine-tuning criterion $\Delta_{EW}$ in (\ref{DeltaEW}) can be analyzed in comparison to the Barbieri-Giudice definition \cite{Barbieri:1987fn}:
\begin{equation}
\Delta_{BG}\equiv {\rm Max}(B_{i})/(M_{{\rm Z}}^{2}/2)
\label{DeltaHS}
\end{equation}
in which the coefficients 
\begin{equation}
B_{i}=\left\lbrace
\begin{array}{cc}
B_{H_{d}} = \mid m^{2}_{H_{d}}(\Lambda)/(\tan^{2}\beta -1)\mid, &
B_{\delta H_{d}} = \mid \delta m^{2}_{H_{d}}/(\tan^{2}\beta -1)\mid \\ \\
B_{H_{u}} = \mid m^{2}_{H_{u}}(\Lambda)/(\tan^{2}\beta -1)\mid, &
B_{\delta H_{u}} = \mid \delta m^{2}_{H_{u}}/(\tan^{2}\beta -1)\mid \\ \\
\hspace{-2.2cm}B_{\mu} = \mid -\mu^{2}(\Lambda)\mid , & \hspace{-7.0cm}B_{\delta\mu} = \mid -\delta\mu^{2}\mid
\end{array}
\right.
\label{Bi-HS}
\end{equation}
are evaluated at high-energy scale $\Lambda$. Obviously, the electroweak fine-tuning in Eq. (\ref{DeltaEW}) can also be continued to $\Lambda$ scale
via renormalization group running and inclusion of the threshold corrections \cite{Baer:2012mv}.  

Note that in the calculation of $\Delta_{BG}$, $\delta m_{H_{u,d}}^{2}$ are considered separately in contrast to those in Eq. (\ref{DeltaEW}). In fact, $\Delta_{BG}$ is calculated in terms of high scale parameters such as  $m_{H_{u,d}}^{2}(\Lambda)$ and $\mu(\Lambda)$ where $\Lambda$ denotes the highest energy scale up to which the model under concern is a valid effetive field theory. In this approach,
$\Delta_{BG}$ contains information on possible high scale origin of the fine-tuning. Gravity mediated supersymmetric theories such as mSUGRA are, in general, assumed to be valid up to the GUT scale, and hence $\Delta_{BG}$ is calculated with terms that are normalized at $\Lambda = M_{{\rm GUT}}$. In such  models, $B_{\delta H_{u}}$ becomes dominant because of large logarithms.  Also $m_{H_{u}}^{2}$ needs a significant contribution, $\delta m_{H_{u}}^{2}$, since it is required to evolve to negatives values from its high scale values as required by radiative electroweak symmetry breaking (REWSB) \cite{Murayama:1995fn}. 
 
The essential observation is that light stop can hide in the top signal in the region with $m_{\tilde{t}_1}\lesssim 200$ GeV. The reported top results can thus contain stop signal within the exclusion limits. In the present paper, our goal is to determine SUSY parameter regions accomodating light stops. In doing this, we consider constrained MSSM (CMSSM) and SUSY GUTs, and investigate their ``light stop regions'' by taking into account bounds from 125 {\rm GeV} Higgs boson, $B$-physics, and cold dark matter. We also give the results for naturalness in terms of $\Delta_{EW}$ and $\Delta_{BG}$ in a general fashion. More detailed studies of naturalness can be found in \cite{Hall:2011aa}. 
 
Outline of the paper is as follows. In Sec. II we describe scanning procedure and various experimental bounds to be imposed. The Sec. III contains our results for CMSSM. The Sec. IV is devoted to nonuniversal Higgs mass (NUHM1) models. The Sec. V discusses the CMSSM with gravitino LSP. The Sec. VI deals with nonuniversal gaugino masses and negative $\mu$. Finally, we conclude in Sec. V.

\section{Scanning Procedure and Experimental Constraints}
\label{sec:scan}

We employ the ISAJET 7.84 package \cite{Paige:2003mg} to perform random scans over the parameter space. In this package, the weak scale values of gauge and third generation Yukawa couplings are evolved to $M_{{\rm GUT}}$ via renormalization group equations (RGEs) in the $\overline{DR}$ regularization scheme. We do not strictly enforce the gauge unification condition $g_{1}=g_{2}=g_{3}$ at $M_{{\rm GUT}}$, since a few percent deviation from unification can be assigned to unknown GUT-scale threshold corrections \cite{Hisano:1992jj}. With the boundary conditions given at $M_{{\rm GUT}}$, all the soft supersymmetry breaking (SSB) parameters, along with the gauge and Yukawa couplings, are evolved back to the weak scale $M_{{\rm Z}}$. 

In RG-evolution of Yukawa couplings the SUSY threshold corrections \cite{Pierce:1996zz} are taken into account at the common scale $M_{{\rm SUSY}}=\sqrt{m_{\tilde{t}_{L}}m_{\tilde{t}_{R}}}$. The entire parameter set is iteratively run between $M_{{\rm Z}}$ and $M_{{\rm GUT}}$ using the full 2-loop RGEs until a stable solution is obtained. To better account for leading-log corrections, 1-loop step beta functions are adopted for gauge and Yukawa couplings, and the SSB parameters $m_{i}$ are extracted from RGEs at appropriate scales $m_{i}=m_{i}(m_{i})$. The RGE-improved 1-loop effective potential is minimized at an optimized scale $M_{{\rm SUSY}}$, which effectively accounts for the leading 2-loop corrections. Full 1-loop radiative corrections are incorporated for all sparticle masses. 

We perform Markov-chain Monte Carlo (MCMC) scans over the parameter spaces of the models we analyze. We also set $\mu > 0$ and $m_{t} = 173.3$ GeV \cite{ATLAS:2014wva,Gogoladze:2014hca}. Note that our results are not too sensitive to variation in the value of $m_{t}$ within $1\sigma-2\sigma$ \cite{bartol2}.

We employ the Metropolis-Hasting algorithm as described in \cite{Belanger:2009ti}, and require all points to satisfy radiative electroweak symmetry breaking (REWSB). The REWSB requirement is a crucial  theoretical constraint on the parameter space \cite{Ibanez:1982fr}. After collecting data, we subsequently impose the mass bounds \cite{Nakamura:2010zzi} and B-physics constraints, stop-top degeneracy band ($\Delta m_{\tilde{t}t} \leq 30$ GeV), and the WMAP bound on the relic density of neutralino LSP. The B-physics observables and relic density of neutralino LSP are calculated by use of IsaTools \cite{bsg,bmm}. The experimental constraints imposed in our data can be summarized as follows:

\begin{eqnarray} 
m_h  = \left( 123-127\right)~{\rm GeV}~~&\cite{ATLAS, CMS} \nonumber & \\
m_{\tilde{g}} \geq 1 {\rm TeV} ~~&\cite{TheATLAScollaboration:2013fha}& \nonumber 
\\
0.8\times 10^{-9} \leq{\rm BR}(B_s \rightarrow \mu^+ \mu^-) 
  \leq 6.2 \times10^{-9} \;(2\sigma)~~&\cite{Aaij:2012nna}& \nonumber
\\ 
2.99 \times 10^{-4} \leq 
  {\rm BR}(b \rightarrow s \gamma) 
  \leq 3.87 \times 10^{-4} \; (2\sigma)~~&\cite{Amhis:2012bh}&  
\\
0.15 \leq \frac{
 {\rm BR}(B_u\rightarrow\tau \nu_{\tau})_{\rm MSSM}}
 {{\rm BR}(B_u\rightarrow \tau \nu_{\tau})_{\rm SM}}
        \leq 2.41 \; (3\sigma)~~&\cite{Asner:2010qj}&  \nonumber
\\
 0.0913 \leq \Omega_{\rm CDM}h^2 (\rm WMAP9) \leq 0.1363   \; (5\sigma)~~&\cite{WMAP9}& \nonumber
%\\ 
% 2.1 \times 10^{-10} \leq \Delta a_{\mu} 
%  \leq 50.1 \times 10^{-10} \; (3\sigma)~~&\cite{BNL}&
\label{constraints}
\end{eqnarray} 

We display the mass bounds on the Higgs boson and gluino, because they have changed since the LEP era. We take the lower bound on gluino mass as $m_{\tilde{g}} \geq 1$ TeV all over the parameter space since the exclusion curve excludes the gluino of mass less than 1 TeV for the LSP of mass less than 300 GeV. Besides these constraints, we require our solution to do no worse than the SM in prediction of the muon 
anomalous magnetic moment (muon $g-2$).

\section{CMSSM}
\label{sec:CMSSM}
%\label{sec:res}
In this section, we study the CMSSM parameter space to determine under what conditions it can accommodate stop-top degeneracy. In doing this we 
scan their parameter spaces to determine viable regions and also compute naturalness of those regions. 

%\subsection{CMSSM}
%\label{subsec:CMSSM}
\begin{figure}\hspace{-0.5cm} 
\includegraphics[scale=0.6]{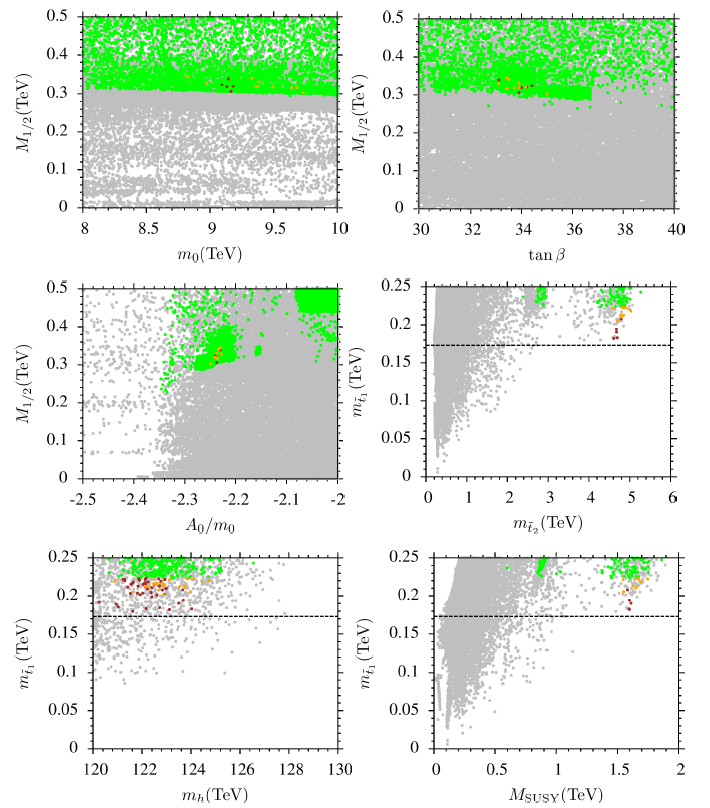}
\caption{Plots in $M_{1/2}-m_{0}$, $M_{1/2}-\tan\beta$, $M_{1/2}-A_{0}/m_{0}$, $m_{\tilde{t}_{1}}-m_{\tilde{t}_{2}}$, $m_{\tilde{t}_{1}}-M_{{\rm SUSY}}$, $m_{\tilde{t}_{1}}-m_{h}$ planes for CMSSM with neutralino LSP. All points are consistent with REWSB and neutralino LSP. Green points satisfy mass bounds and B-physics bounds mentioned in Section \ref{sec:scan}. The red points within the green are the ones that satisfy $\Delta m_{\tilde{t}t} \leq 50$ GeV where $\Delta m_{\tilde{t}t} \equiv m_{\tilde{t}_{1}}-m_{t}$. The orange points are a subset of red points and satisfy the WMAP bound on relic abundance of neutralino within $5\sigma$. The dashed line corresponds to top quark mass $m_{t}=173.3$ {\rm GeV}.}
\label{fundamentals}
\end{figure}

\begin{figure}\hspace{-2.0cm}
\includegraphics[scale=0.65]{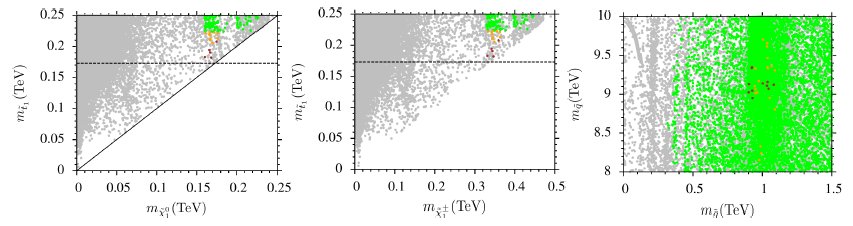}
\caption{The sparticle spectrum in $m_{\tilde{t}_{1}}-m_{\tilde{\chi}_{1}^{0}}$, $m_{\tilde{t}_{1}}-m_{\tilde{\chi}_{1}^{\pm}}$, and $m_{\tilde{q}}-m_{\tilde{g}}$ planes for CMSSM with neutralino LSP. 
The color coding is the same as in Figure \ref{fundamentals} except the lower mass bound on gluino is not applied in the $m_{\tilde{q}}-m_{\tilde{g}}$ panel.}
\label{spect}
\end{figure}

%\begin{figure}[htp!]
%\subfigure{\includegraphics[scale=1.1]{CMSSM_mhf_m0.png}}
%\subfigure{\includegraphics[scale=1.1]{CMSSM_mhf_tanb.png}}
%\subfigure{\includegraphics[scale=1.1]{CMSSM_mhf_a0m0.png}}
%\subfigure{\includegraphics[scale=1.1]{CMSSM_stop1_stop2.png}}
%\subfigure{\includegraphics[scale=1.1]{CMSSM_stop1_mh0.png}}
%\subfigure{\includegraphics[scale=1.1]{CMSSM_stop1_MSUSY.png}}
%\caption{Plots in $M_{1/2}-m_{0}$, $M_{1/2}-\tan\beta$, $M_{1/2}-A_{0}/m_{0}$, $m_{\tilde{t}_{1}}-m_{\tilde{t}_{2}}$, $m_{\tilde{t}_{1}}-M_{{\rm SUSY}}$, $m_{\tilde{t}_{1}}-m_{h}$ planes. All points are consistent with REWSB and neutralino LSP. Green points satisfy mass bounds and B-physics bounds mentioned in Section \ref{sec:scan}. The red points within the green are the ones that satisfy $\Delta m_{\tilde{t}t} \leq 50$ GeV where $\Delta m_{\tilde{t}t} \equiv m_{\tilde{t}_{1}}-m_{t}$. The orange points are a subset of red points and satisfy the WMAP bound on relic abundance of neutralino within $5\sigma$. The dashed line corresponds to top quark mass $m_{t}=173.3$ {\rm GeV}.}
%\label{fundamentals}
%\end{figure}
 
In this section, we scan CMSSM parameter space within the following ranges:
\begin{eqnarray}
 0 \leq m_{0} \leq 20~{\rm TeV} \nonumber \\
  0 \leq M_{1/2} \leq 5~{\rm TeV} \nonumber \\
  -3 \leq A_{0}/m_{0} \leq 3 \\
  2 \leq \tan\beta \leq 60 \nonumber
  \label{CMMS_param}
 \end{eqnarray} 
where $m_{0}$ is the universal SSB mass term for all scalars including $H_{u}$ and $H_{d}$ of the MSSM, and $M_{1/2}$ is the universal SSB gaugino mass term. The $A_{0}$ is SSB trilinear scalar interaction term, and $\tan\beta$ is the ratio of vacuum expectation values of the MSSM Higgs fields.

Figure \ref{fundamentals} displays the plots in $M_{1/2}-m_{0}$, $M_{1/2}-\tan\beta$, $M_{1/2}-A_{0}/m_{0}$, $m_{\tilde{t}_{1}}-m_{\tilde{t}_{2}}$, $m_{\tilde{t}_{1}}-M_{{\rm SUSY}}$, $m_{\tilde{t}_{1}}-m_{h}$ planes. All points are consistent with REWSB and neutralino LSP. Green points satisfy mass bounds and B-physics mentioned in Section \ref{sec:scan}. Red points form a subset of green and they satisfy $\Delta m_{\tilde{t}t} \leq 50$ GeV where $\Delta m_{\tilde{t}t} \equiv m_{\tilde{t}_{1}}-m_{t}$. Orange points are a subset of red points satisfying the WMAP bound on relic abundance of neutralino within $5\sigma$. The dashed line corresponds to top quark mass $m_{t}=173.3$ GeV. Results displayed in $M_{1/2}-m_{0}$, $M_{1/2}-\tan\beta$, $M_{1/2}-A_{0}/m_{0}$ panels show that the degeneracy between top quark and its supersymmetric partner can be realized in a very small region with $m_{0} \sim 9$ TeV, $M_{1/2} \sim 300-350$ GeV, $\tan\beta \sim 34$, and $A_{0}/m_{0} \sim -2.2$.  The large mixing seen in $M_{1/2}-A_{0}/m_{0}$ plane leads to a huge mass difference between the two stop mass eigenstates, $\tilde{t}_1$ and $\tilde{t}_2$. It is seen from $m_{\tilde{t}_{1}}-m_{\tilde{t}_{2}}$ plane that the heavier stop eigenstate has a mass about $4-5$ TeV, while the lighter one stays close to the top quark. This large mixing is also required to satisfy the constraint of 125 GeV Higgs boson. The $m_{\tilde{t}_{1}}-m_{h}$ shows that it is possible to find solutions with the Higgs boson of mass about 124 GeV. Note that the Higgs mass is calculated in ISAJET with an approximate error of about 3 GeV arising from theoretical uncertainties in calculation of the minimum of the scalar potential, and experimental uncertainties in the values of $m_{t}$ and $\alpha_{s}$. From the $m_{\tilde{t}_{1}}-M_{{\rm SUSY}}$ plot, the SUSY decoupling scale reads $M_{{\rm SUSY}} \simeq 1.5$ TeV, and it provides the desired solution for gauge hierarchy problem within the TeV-scale SUSY. It is true that the orange points form a tiny subset; however, once model parameters are fixed by them the resulting model gives a viable SUSY description because stop is sitting on top quark. 

%\begin{figure}[htp!]\hspace{-2.0cm}
%\subfigure{\includegraphics[scale=0.9]{CMSSM_stop1_neut.png}}%
%\subfigure{\hspace{0.01cm}\includegraphics[scale=0.9]{CMSSM_stop1_char.png}}%
%\subfigure{\hspace{0.05cm}\includegraphics[scale=0.9]{CMSSM_uL_glu.png}}
%\caption{The sparticle spectrum in $m_{\tilde{t}_{1}}-m_{\tilde{\chi}_{1}^{0}}$, $m_{\tilde{t}_{1}}-m_{\tilde{\chi}_{1}^{\pm}}$, and $m_{\tilde{q}}-m_{\tilde{g}}$ planes. The color coding is the same as in Figure \ref{fundamentals} except the lower mass bound on gluino is not applied in the $m_{\tilde{q}}-m_{\tilde{g}}$ panel.}
%\label{spect}
%\end{figure}

Spectrum for other SUSY particles is represented in $m_{\tilde{t}_{1}}-m_{\tilde{\chi}_{1}^{0}}$, $m_{\tilde{t}_{1}}-m_{\tilde{\chi}_{1}^{\pm}}$, and $m_{\tilde{q}}-m_{\tilde{g}}$ planes of Figure \ref{spect}. The color coding is the same as Figure \ref{fundamentals} except the lower mass bound on gluino is not applied in the $m_{\tilde{q}}-m_{\tilde{g}}$ panel. In the light stop region it is also found that the lightest stop is almost degenerate with LSP neutralino of mass about 160 GeV, as seen from the $m_{\tilde{t}_{1}}-m_{\tilde{\chi}_{1}^{0}}$ panel. The solutions with $m_{\tilde{\chi}_{1}^{0}} \lesssim 150$ GeV are not consistent with the experimental constraints mentioned in Section \ref{sec:scan}. Similarly the lightest chargino is of mass about 300 GeV in the same region. The $m_{\tilde{q}}-m_{\tilde{g}}$ plane reveals important results for the color sector obtained in the light stop region. The squarks of the first generations are found to be of mass about 9 TeV that is beyond the exclusion limit. On the other hand, the gluino weighs slightly above 1 TeV.
\begin{figure}[t!]
\centering
\subfigure{\includegraphics[scale=1.1]{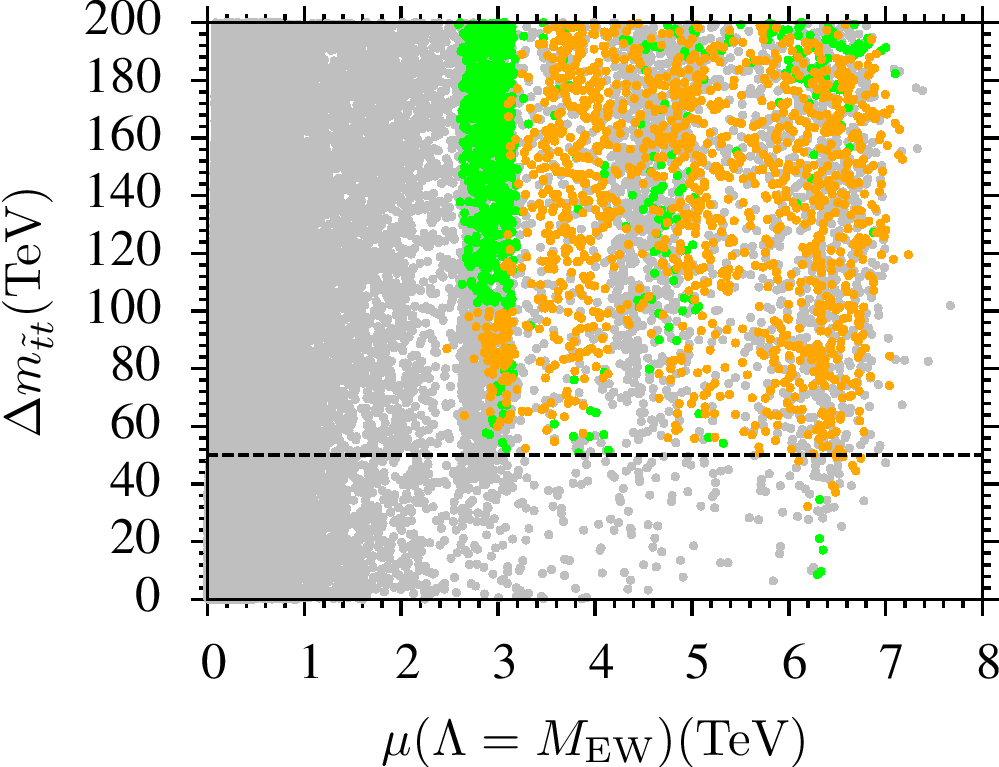}}%
\subfigure{\includegraphics[scale=1.1]{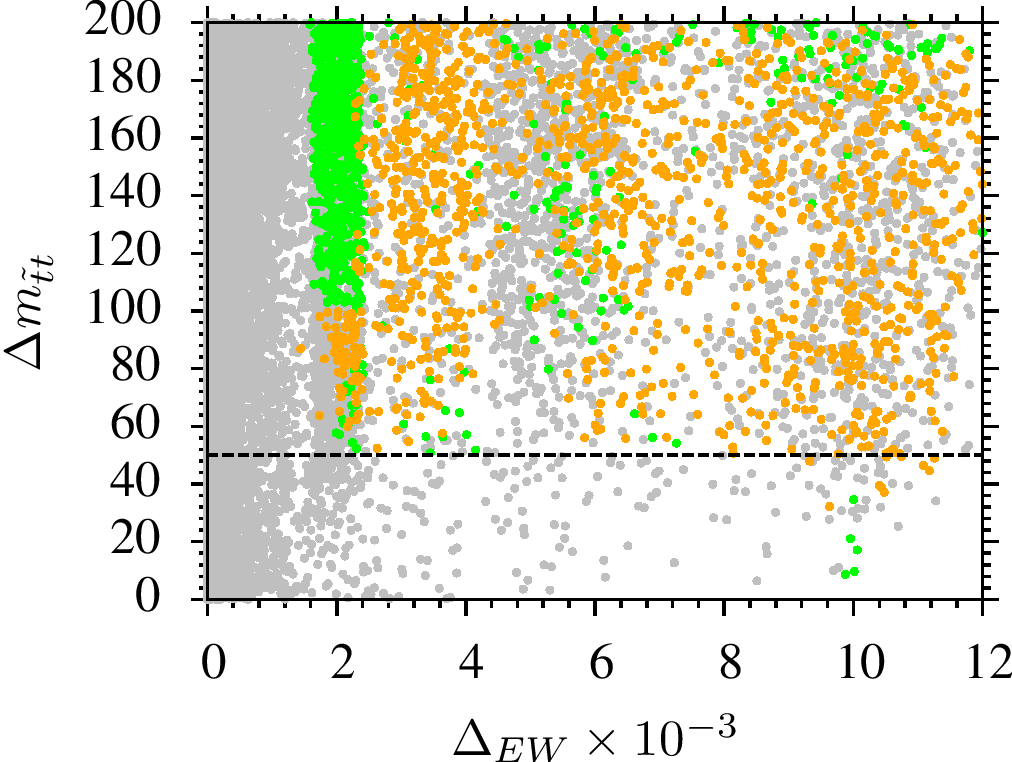}}
\subfigure{\includegraphics[scale=1.1]{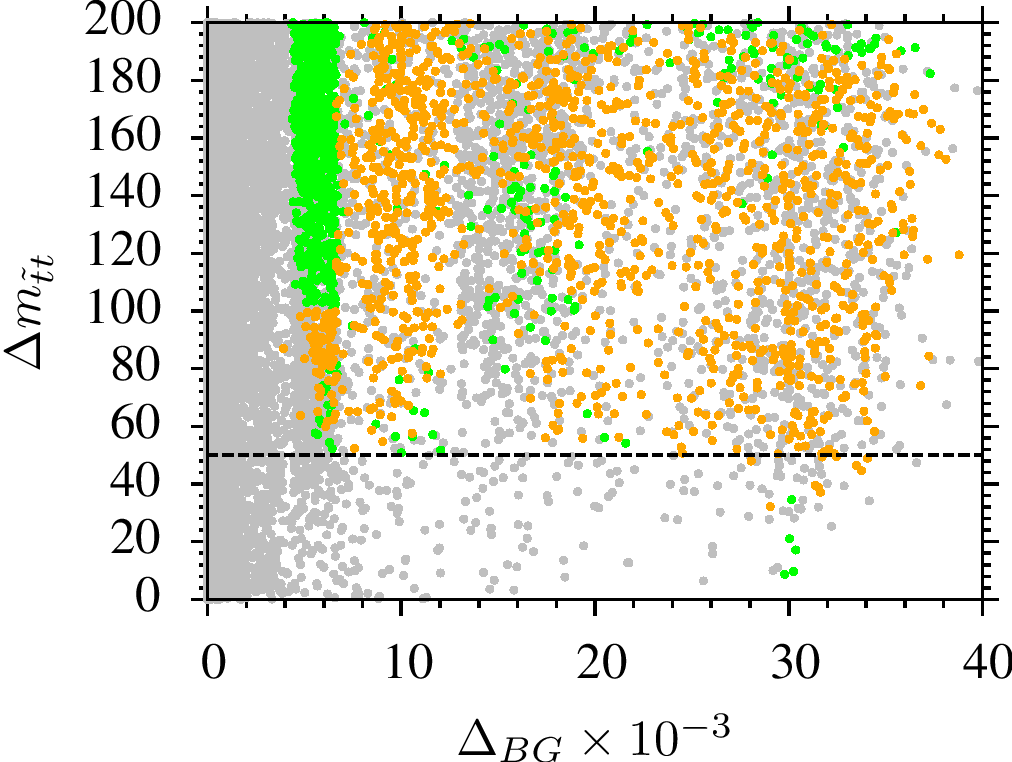}}%
\subfigure{\includegraphics[scale=1.1]{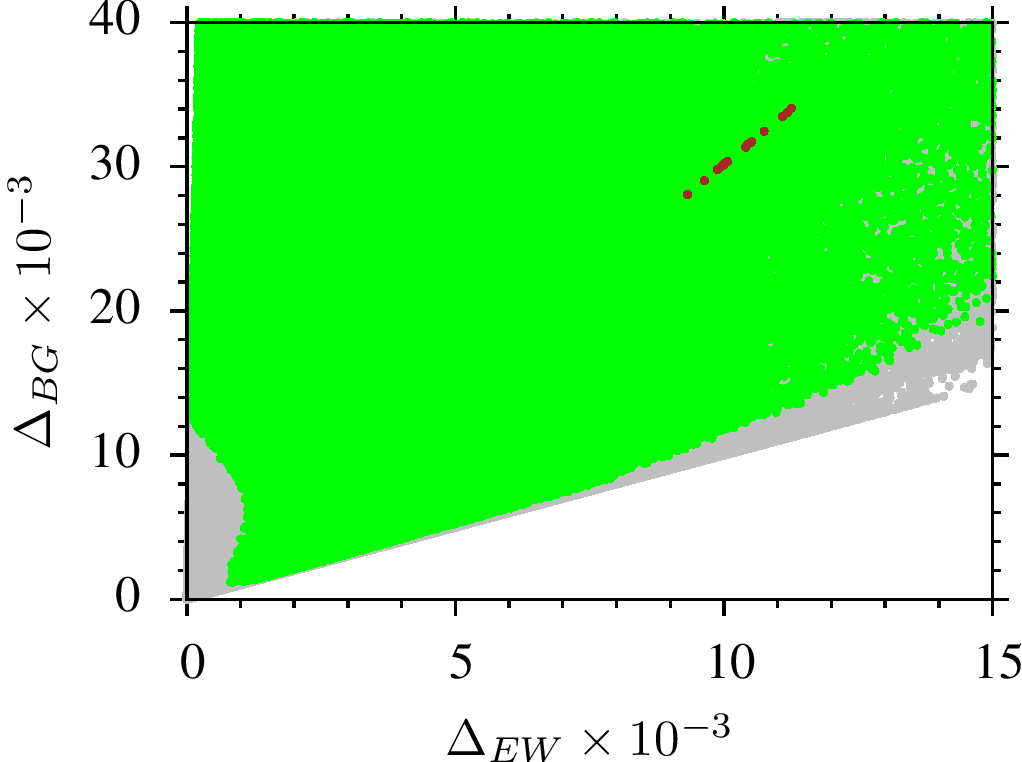}}
\caption{Fine-tuning plots in $\Delta m_{\tilde{t}t}-\mu$, $\Delta m_{\tilde{t}t}-\Delta_{EW}$, $\Delta m_{\tilde{t}t}-\Delta_{BG}$, and $\Delta_{BG}-\Delta_{EW}$ planes for CMSSM with neutralino LSP. The color coding is the same as Figure \ref{fundamentals} except that $\Delta m_{\tilde{t}t} \leq 50$ is not applied in the $\Delta m_{\tilde{t}t}-\mu$, $\Delta m_{\tilde{t}t}-\Delta_{EW}$, $\Delta m_{\tilde{t}t}-\Delta_{BG}$ panels, a guide line is rather used to indicate $\Delta m_{\tilde{t}t}=50$. }
\label{CMSSM_finetuning}
\end{figure}

%\begin{figure}[h!]  
%\vspace{-2.5cm}
%\subfigure{\includegraphics[scale=1.1]{CMSSM_deltamt_muLOW.png}}
%\subfigure{\includegraphics[scale=1.1]{CMSSM_deltamt_DeltaEW.png}}
%\subfigure{\includegraphics[scale=1.1]{CMSSM_deltamt_DeltaHS.png}}
%\subfigure{\includegraphics[scale=1.1]{CMSSM_DeltaEW_DeltaHS.png}}
%\caption{Fine-tuning plots in $\Delta m_{\tilde{t}t}-\mu$, $\Delta m_{\tilde{t}t}-\Delta_{EW}$, $\Delta m_{\tilde{t}t}-\Delta_{BG}$, and $\Delta_{BG}-\Delta_{EW}$ planes. The color coding is the same as Figure \ref{fundamentals} except that $\Delta m_{\tilde{t}t} \leq 50$ is not applied in the $\Delta m_{\tilde{t}t}-\mu$, $\Delta m_{\tilde{t}t}-\Delta_{EW}$, $\Delta m_{\tilde{t}t}-\Delta_{BG}$ panels, a guide line is rather used to indicate $\Delta m_{\tilde{t}t}=50$. }
%\label{CMSSM_finetuning}
%\end{figure}

Figure \ref{CMSSM_finetuning} shows the results for the fine-tuning calculated for the light stop region in $\Delta m_{\tilde{t}t}-\mu$, $\Delta m_{\tilde{t}t}-\Delta_{EW}$, $\Delta m_{\tilde{t}t}-\Delta_{BG}$, and $\Delta_{BG}-\Delta_{EW}$ planes. The color coding is the same as Figure \ref{fundamentals} except that $\Delta m_{\tilde{t}t} \leq 50$ is not applied in the $\Delta m_{\tilde{t}t}-\mu$, $\Delta m_{\tilde{t}t}-\Delta_{EW}$, $\Delta m_{\tilde{t}t}-\Delta_{BG}$ panels, a guide line is rather used to indicate $\Delta m_{\tilde{t}t}=50$. The fact that this region needs a large mixing between the stops results in a large SSB trilinear $A_{t}-$term, and it leads to a significant fine-tuning as seen from the plots of Figure \ref{CMSSM_finetuning}. The $\Delta m_{\tilde{t}t}-\mu$ plane shows that $\mu(\Lambda = {\rm weak}) \simeq 3$ TeV for $\Delta m_{\tilde{t}t} \simeq 50$, and it raises up to 6 TeV, if one seeks for $m_{\tilde{t}t} \leq 50$. Similarly $\Delta_{EW}\simeq 2000~(0.05\%)$ and $\Delta_{BG}\simeq 6000~(0.017\%)$ for $\Delta m_{\tilde{t}t}\simeq 50$, while $\Delta_{EW}\simeq 9000~(0.012\%)$ and $\Delta_{BG}\simeq 30000~(0.003\%)$ for $\Delta m_{\tilde{t}t}\leq 50$. The $\Delta_{BG}-\Delta_{EW}$ plane summarizes the results obtained for the fine-tuning. 

%\newpage

\begin{table}[t!] 
\centering
\begin{tabular}{|c|ccc|}
\hline
\hline
                 & Point 1 & Point 2 & Point 3 \\

\hline
$m_{0}$         &9165  & 8975 & 9460  \\
$M_{1/2} $      &305.9 & 328.9 & 308.7 \\
$\tan\beta$     &33.9 & 33.1 & 31.0  \\
$A_0/m_{0}$     &-2.24 & -2.23 & -2.21  \\
$m_t$           &{\color{red}173.3} & {\color{red}173.3} & {\color{red}173.3} \\
\hline
$\mu$          & 6134 &6008 & 6312 \\
$\Delta a_{\mu}$ & $ { 0.50\times 10^{-11} } $ & $ { 0.39\times 10^{-10} } $ & $ { 0.42\times 10^{-11} } $  
\\ 
\hline
$m_h$            & {\color{red} 124.1}& {\color{red} 124} & {\color{red}126.1} \\
$m_H$           & 6922  & 7049 & 8041 \\
$m_A$           & 6878 & 7004 & 7989 \\
$m_{H^{\pm}}$   & 6923 & 7050 & 8042 \\

\hline
$m_{\tilde{\chi}^0_{1,2}}$
                 & \textbf{160.3}, \textbf{328.2}  & \textbf{169.6}, \textbf{345.9}  & \textbf{ 161.1}, \textbf{329.2} \\

$m_{\tilde{\chi}^0_{3,4}}$
                 & 6156, 6156 & 6032, 6032  & 6343, 6344 \\

$m_{\tilde{\chi}^{\pm}_{1,2}}$
                & \textbf{331.2}, 6171 & \textbf{349.1}, 6047 & \textbf{332.3}, 6361 \\

$m_{\tilde{g}}$  & \textbf{1002} & \textbf{1054} & \textbf{1010} \\
\hline $m_{ \tilde{u}_{L,R}}$
                 & 9139, 9172  & 8953, 8980 & 9430, 9460 \\
$m_{\tilde{t}_{1,2}}$
                 &{\color{red} 182.9}, 4696 & {\color{red} 205.4}, 4705 & {\color{red} 213.4}, 5192 \\
\hline $m_{ \tilde{d}_{L,R}}$
                 & 9139, 9172 & 8953, 8985  & 9431, 9465 \\
$m_{\tilde{b}_{1,2}}$
                 & 4803, 6648  & 4813, 6660  & 5311, 7355  \\
\hline
$m_{\tilde{\nu}_{e,\mu}}$
                 &9167 & 8977 & 9462 \\
$m_{\tilde{\nu}_{\tau}}$
                 & 8045 &  7925 & 8482  \\
\hline
$m_{ \tilde{e}_{L,R}}$
                & 9156, 9159 & 8967, 8969  & 9450, 9454 \\
$m_{\tilde{\tau}_{1,2}}$
                & 6855, 8076 & 6819, 7955 & 7483, 8514 \\
\hline

$\sigma_{SI}({\rm pb})$
                & $ { 0.11\times 10^{-11} } $ & $0.11\times 10^{-11}$ & $0.76\times 10^{-12}$ \\

$\sigma_{SD}({\rm pb})$
                &$ { 0.37\times 10^{-10} } $ & $0.40\times 10^{-10}$ & $0.32\times 10^{-10}$ \\

$\Omega h^{2}$      & {\color{blue}0.03} & {\color{red} 0.104} & {\color{blue} 0.18} \\
\hline
$\Delta_{EW}$ & 10014 & 9633 & 10691\\
$\Delta_{BG}$  & 30244 & 29034 & 32248\\
\hline
\hline
\end{tabular}
\caption{Benchmark points with $m_{\tilde{t}_{1}} \simeq m_{t}$ for CMSSM with neutralino LSP. Masses are given in GeV unit. All points are chosen as to be consistent with mass and B-physics constraints. Point 1 displays a solution with the best degeneracy between stop and top quarks. Point 2 depicts a solution consistent with WMAP bound on relic abundance of LSP neutralino within $5\sigma$. Point 3 yields 126 GeV Higgs boson solution with the least mass separation between stop and top quark. These points do not generate or tunnel to a deeper CCB minima.}
\label{benchCMSSM} 
\end{table}

Another important constraint comes from the WMAP9 searches for the dark matter. Mass difference up to $20\%$ between NLSP stop and LSP neutralino is exluded in order to satisfy the WMAP bound on relic abundance of LSP neutralino for $m_{\tilde{\chi}_{1}^{0}} \lesssim 200$ GeV \cite{He:2011tp}. In our results for CMSSM, there is no solution that satisfy both the WMAP bound and $\Delta m_{\tilde{t}t} \leq 30$ GeV. We also illustrate our results for CMSSM in Table \ref{benchCMSSM}. Masses are given in GeV unit. All points are chosen as to be consistent with mass and B-physics constraints. Point 1 displays a solution with the best degeneracy between stop and top quarks. As stated above, relic abundance of LSP neutralino is too low for Point 1 to satisfy the WMAP bound. Point 2 depicts a solution consistent with WMAP bound on relic abundance of LSP neutralino within $5\sigma$. The mass difference between stop and neutralino is about 32 GeV for Point 2. Point 3 yields 126 GeV Higgs boson solution with the least mass separation between stop and top quarks. We checked by using the public code Vevacious \cite{Camargo-Molina:2013qva} that Point 1, Point 2 and Point 3 yield stable vacua.

\section{NUHM1}
\label{sec:NUHM1}
%\label{subsec:NUHM1}

\begin{figure}\hspace{-0.5cm} 
\includegraphics[scale=0.6]{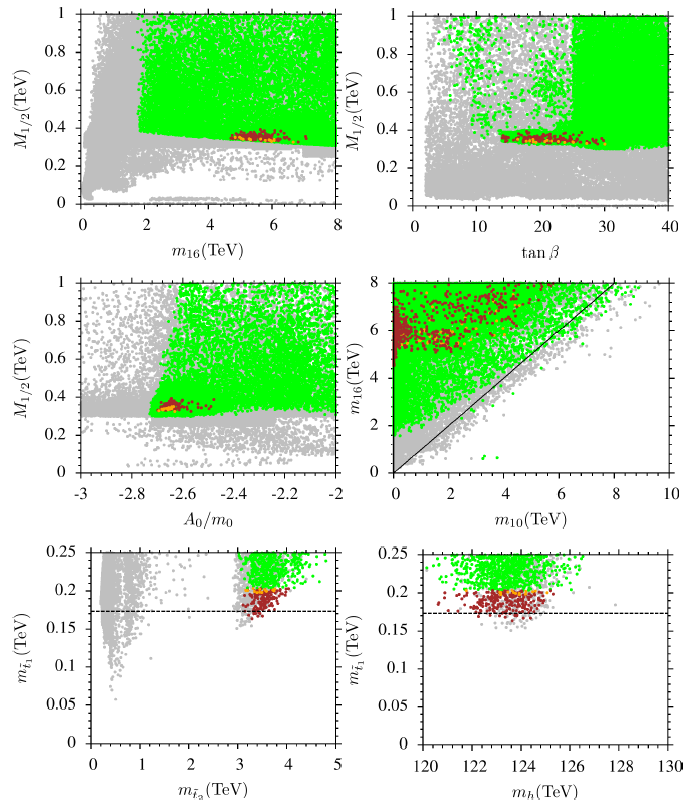}
\caption{Plots in $M_{1/2}-m_{16}$, $M_{1/2}-\tan\beta$, $M_{1/2}-A_{0}/m_{16}$, $m_{16}-m_{10}$, $m_{\tilde{t}_{1}}-m_{\tilde{t}_{2}}$, $m_{\tilde{t}_{1}}-M_{{\rm SUSY}}$, $m_{\tilde{t}_{1}}-m_{h}$ planes for NUHM1 with neutralino LSP. The color coding is the same as Figure \ref{fundamentals} except that $\Delta m_{\tilde{t}t} \leq 30$ GeV is applied in the red region. The solid line in $m_{16}-m_{10}$ plane corresponds to the CMSSM solution for which $m_{16}=m_{10}=m_0$.}
\label{SO10_fundamentals}
\end{figure}

Having found that the CMSSM is severely fine-tuned in producing stop-top degeneracy region, we now start looking for extensios of the CMSSM where fine-tuning is lower. We start our search with the CMSSM with nonunivseral Higgs masses at the unication scale. In other words, we relax the CMSSM setup by separating the SSB mass term for the MSSM Higgs fields from the one for the remaining matter scalars. The results displayed here are obtained from scanning over the following parameter space: 
\begin{eqnarray}
 0 \leq m_{16} \leq 20~{\rm TeV} \nonumber \\
 0 \leq m_{10} \leq 20~{\rm TeV} \nonumber \\
  0 \leq M_{1/2} \leq 5~{\rm TeV}  \\
  -3 \leq A_{0}/m_{0} \leq 3 \nonumber \\
  2 \leq \tan\beta \leq 60 \nonumber
  \label{SO10_param}
 \end{eqnarray} 
where $m_{16}\equiv m_0$ and $m_{10}$ are the SSB mass terms for the matter scalars and the Higgs fields, respectively. All other parameters are the same as in the previous subsection. 

Figure \ref{SO10_fundamentals} displays the scan results in $M_{1/2}-m_{16}$, $M_{1/2}-\tan\beta$, $M_{1/2}-A_{0}/m_{16}$, $m_{16}-m_{10}$, $m_{\tilde{t}_{1}}-m_{\tilde{t}_{2}}$, $m_{\tilde{t}_{1}}-M_{{\rm SUSY}}$ and $m_{\tilde{t}_{1}}-m_{h}$ planes. The color coding is the same as Figure \ref{fundamentals} except that now the red region corresponds to $\Delta m_{\tilde{t}t} \leq 30$ GeV (which is 
not reachable in the CMSSM domain). The solid line in $m_{16}-m_{10}$ plane corresponds to the region where $m_{16}=m_{10}$ (the CMSSM solution). While we have a very narrow range for the SSB gaugino mass term ($M_{1/2} \sim 400$ GeV), compared to the CMSSM regions, the ranges for the other fundamental parameters become slightly wider because now $m_{16} \sim 4.5-7.5$ TeV, $m_{10} \lesssim 5$ TeV, $\tan\beta \sim 14-30$, and $A_{0}/m_{16}\sim$ -$2.7~-$ -$2.4$. Note that the SSB masses for the matter scalars is shifted back to $6$ TeV, while it is strictly $\sim 9$ TeV in CMSSM. As we can see from the $m_{16}-m_{10}$ panel, light Higgs masses at the GUT scale favors light stops at the low scale. On the other hand a large mixing between left and right handed stops can occur because of large $A_{t}$ values, and the heavier stop is found to be of mass about 3-4 TeV. Also, the $m_{\tilde{t}_{1}}-m_{h}$ panel shows that it is possible to find the $SM$-like Higgs boson of mass up to $\sim 126$ GeV. 

\begin{figure}\hspace{-2.0cm}
\includegraphics[scale=0.65]{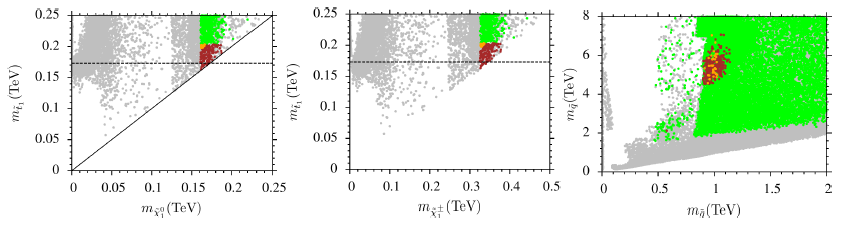}
\caption{SUSY spectrum in $m_{\tilde{t}_{1}}-m_{\tilde{\chi}_{1}^{0}}$, $m_{\tilde{t}_{1}}-m_{\tilde{\chi}_{1}^{\pm}}$, and $m_{\tilde{q}}-m_{\tilde{g}}$ planes for NUHM1 with neutralino LSP. The color coding is the same as Figure \ref{SO10_fundamentals} except the lower mass bound on gluino is not applied in the $m_{\tilde{q}}-m_{\tilde{g}}$ panel.}
\label{SO10_spect}
\end{figure}

We depict the results for the SUSY spectrum in $m_{\tilde{t}_{1}}-m_{\tilde{\chi}_{1}^{0}}$, $m_{\tilde{t}_{1}}-m_{\tilde{\chi}_{1}^{\pm}}$, and $m_{\tilde{q}}-m_{\tilde{g}}$ planes of Figure \ref{SO10_spect}. The color coding is the same as in Figure \ref{SO10_fundamentals} except that the lower mass bound on gluino is not applied in the $m_{\tilde{q}}-m_{\tilde{g}}$ panel. Despite similar results to those obtained for CMSSM, in the NUHM1 framework it is possible to have solution with $m_{\tilde{t}_{1}} \lesssim m_{t}$. We have an extreme degeneracy between stop and neutralino, and hence neutralino relic abundance is too low as stated in the previous section. The $m_{\tilde{t}_{1}}-m_{\tilde{\chi}_{1}^{\pm}}$ panel indicates the chargino mass range $\sim 300-400$ GeV. Squarks of the first generations have mass $\gtrsim 4$ TeV, while the gluinos are found to be of mass $1-1.2$ TeV, as seen in the $m_{\tilde{q}}-m_{\tilde{g}}$ plane.

\begin{figure}[h!]  
\includegraphics[scale=0.6]{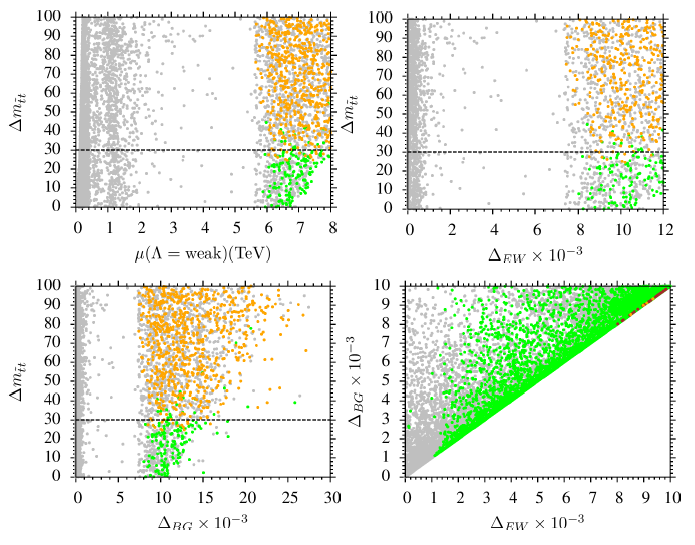}
\caption{Fine-tuning plots in $\Delta m_{\tilde{t}t}-\mu$, $\Delta m_{\tilde{t}t}-\Delta_{EW}$, $\Delta m_{\tilde{t}t}-\Delta_{BG}$, and $\Delta_{BG}-\Delta_{EW}$ planes for NUHM1 with neutralino LSP. The color coding is the same as Figure \ref{CMSSM_finetuning}. }
\label{SO10_finetuning}
\end{figure}

We represent the results for the fine-tuning in $\Delta m_{\tilde{t}t}-\mu$, $\Delta m_{\tilde{t}t}-\Delta_{EW}$, $\Delta m_{\tilde{t}t}-\Delta_{BG}$, and $\Delta_{BG}-\Delta_{EW}$ planes of Figure \ref{SO10_finetuning}. The color coding is the same as Figure \ref{CMSSM_finetuning}. The results are similar to those obtained for CMSSM. $\mu(\Lambda = {\rm weak}) \sim 6-8$ TeV and $\Delta_{EW}\sim 8000-12000~(0.012\% - 0.008\%)$ for $\Delta m_{\tilde{t}t}\leq 30$, while we observe an improvement in $\Delta_{BG}$ as $\sim 8000-16000~(0.012\% - 0.06\%)$ in compare to CMSSM with neutralino LSP. 

\begin{table}[h!] \hspace{-1.4cm}
\centering
\begin{tabular}{|c|cccc|}

\hline
\hline
                 & Point 1 &  Point 2 & Point 3 & Point 4\\

\hline
$m_{0}$          & 5051 & 5641 & 6267 & 5796\\
$m_{10}$        & 45.3 &  21.9 & 190  & 76.6\\
$M_{1/2} $       & 359.4 & 393.1 & 341.7 & 335\\
$\tan\beta$     & 18.0 & 17.9 & 20.6 & 19.4\\
$A_0/m_{0}$    & -2.64 & -2.64 & -2.67 & -2.66\\
$m_t$            & {\color{red}173.3} & {\color{red} 173.3} & {\color{red} 173.3} & {\color{red} 173.3}\\
\hline
$\mu$          & 6168 & 6887 & 7612 & 7050\\
$\Delta a_{\mu}$  & $ { 0.77\times 10^{-11} }$ & $0.60\times 10^{-11}$  & $0.49\times 10^{-11}$ & $0.56\times 10^{-11}$\\ 
\hline
$m_h$            &  \textbf{123.4} &  \textbf{123.8} & \textbf{125.3} & \textbf{125.1}\\
$m_H$           & 5591 & 6258 & 6647 & 6266\\
$m_A$           & 5556 & 6218 & 6604  & 6225\\
$m_{H^{\pm}}$    & 5593 & 6260 & 6648 & 6267\\

\hline
$m_{\tilde{\chi}^0_{1,2}}$
                   & \textbf{ 170.4}, \textbf{340.6}  & \textbf{187.4}, \textbf{374.1} & \textbf{166.9}, \textbf{335.8} & \textbf{162.1}, \textbf{325.9}\\

$m_{\tilde{\chi}^0_{3,4}}$
                 & 6159, 6159  & 6877, 6878 & 7600, 7600 & 7040, 7040\\

$m_{\tilde{\chi}^{\pm}_{1,2}}$
                 & \textbf{343}, 6165 & \textbf{376.7}, 6884 & \textbf{338.1}, 7606 & \textbf{328.1}, 7046\\

$m_{\tilde{g}}$   & \textbf{1047} & \textbf{1189} & \textbf{1054} & 1020\\
\hline $m_{ \tilde{u}_{L,R}}$
                 & 5071, 5083 & 5662, 5676 & 6269, 6286 & 5801, 5816\\
$m_{\tilde{t}_{1,2}}$
                  & {\color{red} 173.3}, 3313 & {\color{red} 189.6}, 3702 & {\color{red} 198.5}, 3985 & {\color{red} 200.2}, 3735\\
\hline $m_{ \tilde{d}_{L,R}}$
                  & 5072, 5083  & 5663, 5676 & 6270, 6286 & 5802, 5817\\
$m_{\tilde{b}_{1,2}}$
                 & 3346, 4694  & 3743, 5250  & 4026, 5659 & 3773, 5300\\
\hline
$m_{\tilde{\nu}_{e,\mu}}$
                  & 5059 & 5650 & 6275 & 5804\\
$m_{\tilde{\nu}_{\tau}}$
                  &  4842 & 5412 & 5929 & 5517\\
\hline
$m_{ \tilde{e}_{L,R}}$
                 & 5053, 5048  & 5643, 5637 & 6267, 6262 & 5796, 5792\\
$m_{\tilde{\tau}_{1,2}}$
                 & 4621, 4847 & 5170, 5417 & 5581, 5937 & 5229, 5524\\
\hline

$\sigma_{SI}({\rm pb})$
                & $0.32\times 10^{-11}$ & $0.20\times 10^{-11}$ & $0.14\times 10^{-11}$ & $0.11\times 10^{-11}$\\

$\sigma_{SD}({\rm pb})$
                & $0.61\times 10^{-11}$ & $0.41\times 10^{-11}$ & $0.27\times 10^{-11}$ & $0.37\times 10^{-11}$\\

$\Omega h^{2}$       & $0.14\times 10^{-3}$ & $0.18\times 10^{-3}$ & 0.07 & \textbf{0.11}\\
\hline
$\Delta_{EW}$ & 9206 & 11478 & 14044 & 12037\\
$\Delta_{BG}$ & 9207 & 11478 & 14052 & 12039\\

\hline
\hline
\end{tabular}
\caption{Benchmark points with $m_{\tilde{t}_{1}} \simeq m_{t}$ in NUHM1 with neutralino LSP. Masses are given in GeV unit, and all points are chosen as to be consistent with mass and B-physics constraints. Point 1 depicts a solution with exact degeneracy between stop and top quarks. Point 2 and Point 3 display the heaviest gluino and the heaviest Higgs boson solutions respectively. Point 4 represents a solution that is consistent with WMAP bound on relic abundance of neutralino LSP.} 
\label{SO10_bench}
\end{table}

Finally, we represent some bencmark points that exemplify our results obtained for NUHM1 in Table \ref{SO10_bench}. Masses are given in GeV unit, and all points are chosen as to be consistent with mass and B-physics constraints. Point 1 depicts a solution with exact degeneracy between stop and top quarks. Point 2 and Point 3 display the heaviest gluino and the heaviest Higgs boson solutions respectively. Point 4 represents a solution that is consistent with WMAP bound on relic abundance of neutralino LSP. The main lesson from this section is that nonuniversality in Higgs mass parameters typically reduce fine-tuning from $10^{4}$ to $10^{3}$ level. This is good but certainly insufficient for having a sensible parameter domain for stop-top degeneracy. 

%\newpage

\section{Gravitino LSP}
\label{gravLSP}

In Sec. 3 and 4, we have considered and revealed our results for solutions with $\mid m_{\tilde{t}}-m_{t} \mid \leq 30$ GeV in CMSSM and NUHM1. In scanning the parameter spaces of these models, we have accepted only the solutions that are compatible with neutralino LSP, and found that the mass difference between NLSP stop and LSP neutralino up to $20\%$ is excluded by the WMAP bound. However; there are some other possible dark matter candidates in the MSSM such as sneutrino and gravitino. In the case of left-handed sneutrino LSP, the sneutrinos lighter than 25 GeV have been excluded by  LEP searches for Z-decays, while those heavier than 25 GeV would have already been observed in direct detection experiments \cite{Baer:2006rs}. Then, one may consider the gravitino LSP case as an alternative to neutralino LSP cases. Although its interactions are too weak to be detected at the LHC, the gravitino LSP is strongly constrained by Big Bang nucleosynthesis (BBN) \cite{Kawasaki:2008qe}, observed abundances of primordial light elements such as D, He, and Li \cite{Pradler:2007is}, and by the LHC and WMAP constraints. Especially, BBN stringently constrains the gravitino LSP. Also, since gravitino relic abundance receives contributions from decays of next to lightest supersymmetric particle (NLSP), the properties of NLSP become important and constrained by the cosmological observations. Each kind of NLSP has its own phenomenology, and many possibilities have been studied such as neutralino, stau, sneutrino, and stop \cite{Ellis:2008as, DiazCruz:2007fc}.

The gravitino mass is proportional to the SUSY breaking scale 
\begin{equation}
\label{gravitino mass}
m_{3/2} \sim \frac{\left\langle F \right\rangle}{\sqrt{3}M_{{\rm Pl}}}
\end{equation}
where $m_{3/2}$ is the gravitino mass, $F$ is the SUSY breaking scale squared in the hidden sector, and $M_{{\rm P}}$ is the Planck scale. Here $\left\langle F \right\rangle$ depends on types of the messengers that mediate SUSY breaking from the hidden sector to the visible one. In gauge mediated SUSY breaking scenario $\left\langle F \right\rangle \sim 10^{8}-10^{19}$, hence the gravitino mass is in the range of $0.1~{\rm eV}-10~{\rm GeV}$ \cite{Giudice:1998bp}, while in gravity mediated supersymmetry breaking $\left\langle F \right\rangle \sim 10^{21}-10^{22}$ GeV$^2$, and hence $m_{3/2} \sim 100-1000$ GeV \cite{Dimopoulos:1996yq}. Note that it is also possible to vary $m_{3/2}$ as a free parameter of a model by keeping gravitino as the LSP \cite{DiazCruz:2007fc}.  

In the supersymmetric models considered in this paper SUSY is broken via gravity mediation and gravitino is necessarily in the spectrum. In this section, we consider CMSSM in the framework of gravity mediated SUSY breaking with gravitino LSP whose mass is assumed to be $\lesssim 100$ GeV. The presence of gravitino does not affect significantly the remaining sparticle spectrum. The fact that gravitino
itself is the LSP relaxes strong restrictions on CMSSM parameter space. As we adopted in the previous sections, we search for the regions with $m_{\tilde{t}}\simeq m_{t}$. It also leads to NLSP stop with gravitino LSP, and hence it is constrained by the cosmological constraints as well as the gravitino. Indeed, BBN constrains the life time and decays of NLSP \cite{Kawasaki:2008qe}, and the bound on stop is much stronger than other possibilities. Even for $m_{3/2}\sim 10$ GeV the lightest stop should be $\sim 1$ TeV, and it is consistent with the results of \cite{DiazCruz:2007fc} that did not find any solution for NLSP stop in CMSSM. In order to avoid the stringent bounds, we can assume the presence of a slight R-parity violation (RPV) that is consistent with the stops lighter than 200 GeV when $\epsilon \gtrsim 10^{-7}$ \cite{Covi:2014fba}, where $\epsilon$ measures the RPV. In this case, gravitino decays are suppressed by the small R-parity breaking parameter as well as the Planck mass, and hence it forms a viable dark matter even in the case of RPV. 

\begin{figure}[t!]
\includegraphics[scale=0.6]{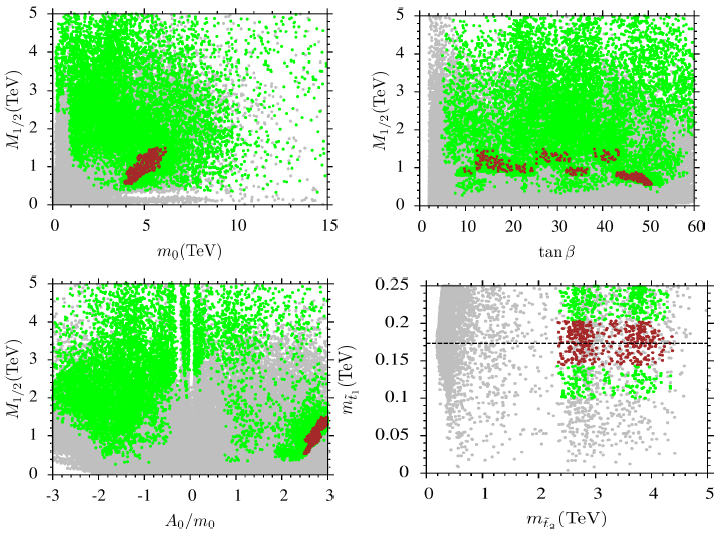}
\caption{Plots in $M_{1/2}-m_{0}$, $M_{1/2}-\tan\beta$, $M_{1/2}-A_{0}/m_{0}$, $m_{\tilde{t}_{1}}-m_{\tilde{t}_{2}}$ planes for CMSSM with gravitino LSP. All points are consistent with REWSB. Green points satisfy mass bounds and B-physics mentioned in Section \ref{sec:scan}. red points form a subset of the green and satisfy $\Delta m_{\tilde{t}t} \leq 30$ GeV where $\Delta m_{\tilde{t}t} \equiv m_{\tilde{t}_{1}}-m_{t}$. The dashed line corresponds to the top quark mass $m_{t}=173.3$ GeV.}
\label{grav_fundamentals}
\end{figure}

With these assumptions, our results in Figure \ref{grav_fundamentals} show that solutions with $m_{\tilde{t}}\sim m_{t}$ can be realized already in the CMSSM. All points are consistent with REWSB. Green points satisfy mass bounds and B-physics mentioned in Section \ref{sec:scan}. Red points from a subset of green and they satisfy $\Delta m_{\tilde{t}t} \leq 30$ GeV where $\Delta m_{\tilde{t}t} \equiv m_{\tilde{t}_{1}}-m_{t}$. The dashed line corresponds to $m_{\tilde{t}_{1}}=m_{t}=173.3$ GeV. The plots show that the region which yield degenerate top and stop is much wider than that found in the CMSSM with neutralino LSP. As shown in the $M_{1/2}-m_{0}$ plane, degeneracy can be realized for $4 \lesssim m_{0} \lesssim 6$ TeV, while $M_{1/2}$ is lifted up to 800 GeV or so. Also, $\tan\beta$ is found to lie in a wide range from $10$ to $50$, as shown in the $M_{1/2}-\tan\beta$ plane. The $M_{1/2}-A_{0}/m_{0}$ plane indicates that one needs a large mixing between left and right-handed stops as in CMSSM with neutralino LSP, but in gravitino LSP case the mixing has opposite sign. With this mixing, the heavier stop is found weigh $\gtrsim 3$ TeV, as expected. 

\begin{figure}[t!]
\includegraphics[scale=0.6]{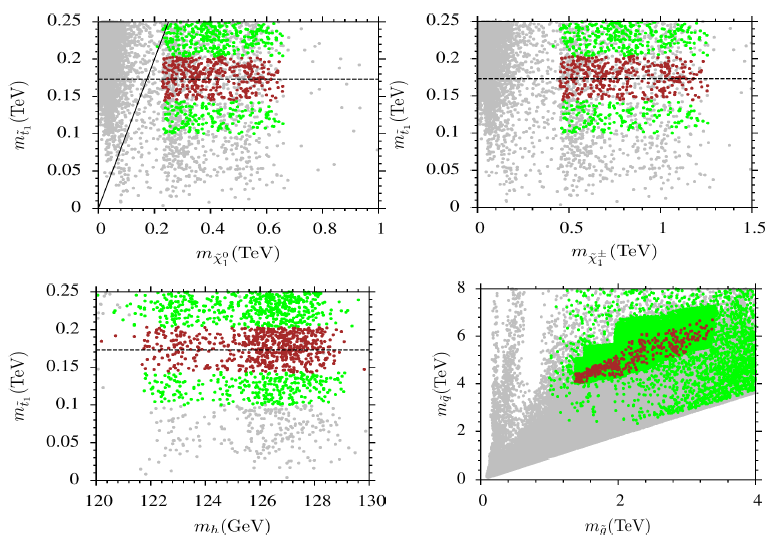}
\caption{Plots in $m_{\tilde{t}_{1}}-m_{\tilde{\chi}_{1}^{0}}$, $m_{\tilde{t}_{1}}-m_{\tilde{\chi}_{1}^{\pm}}$, $m_{\tilde{t}_{1}}-m_{h}$, and $m_{\tilde{q}}-m_{\tilde{g}}$ planes for CMSSM with gravitino LSP. The color coding is the same as Figure \ref{grav_fundamentals}. The solid line in $m_{\tilde{t}_{1}}-m_{\tilde{\chi}_{1}^{0}}$ indicates the region where $m_{\tilde{t}_{1}}=m_{\tilde{\chi}_{1}^{0}}$.}
\label{grav_spec}
\end{figure}

The spectra obtained from our data are represented in $m_{\tilde{t}_{1}}-m_{\tilde{\chi}_{1}^{0}}$, $m_{\tilde{t}_{1}}-m_{\tilde{\chi}_{1}^{\pm}}$, $m_{\tilde{t}_{1}}-m_{h}$, and $m_{\tilde{q}}-m_{\tilde{g}}$ planes of Figure \ref{grav_spec}. The color coding is the same as in Figure \ref{grav_fundamentals}. The solid line in $m_{\tilde{t}_{1}}-m_{\tilde{\chi}_{1}^{0}}$ indicates the region where $m_{\tilde{t}_{1}}=m_{\tilde{\chi}_{1}^{0}}$. Neutralino is found to be heavier ($\sim 200-600$ GeV) than the stop all over the region, and as stated above, this region is excluded, since a charged sparticle becomes LSP unless the gravitino is the LSP. Once we allow the stop to be lighter than neutralino, the chargino is also found to be heavier than those in the CMSSM with neutralino LSP. The $m_{\tilde{t}_{1}}-m_{\tilde{\chi}_{1}^{0}}$ plane shows the range for the lightest chargino mass as $\sim 500-1200$ GeV. Similarly, we have much wider mass range for the Higgs boson mass as $\sim 122 - 130$ GeV. The squarks of the first two generations are found to be lighter ($\sim 4-7$ TeV), while the gluinos are heavy ($\sim 1.5-3.5$ TeV) and they can be tested in LHC14 run. 

\begin{figure}[t!]
\includegraphics[scale=0.6]{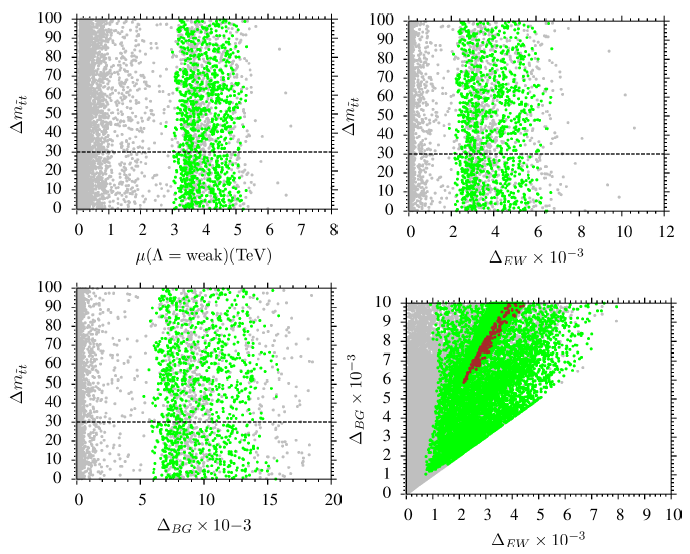}
\caption{Fine-tuning plots in $\Delta m_{\tilde{t}t}-\mu$, $\Delta m_{\tilde{t}t}-\Delta_{EW}$, $\Delta m_{\tilde{t}t}-\Delta_{BG}$, and $\Delta_{BG}-\Delta_{EW}$ planes for CMSSM with gravitino LSP. The color coding is the same as Figure \ref{CMSSM_finetuning}. }
\label{CMSSM_grav_finetuning}
\end{figure}

Figure \ref{CMSSM_grav_finetuning} displays the resulst for the fine-tuning in $\Delta m_{\tilde{t}t}-\mu$, $\Delta m_{\tilde{t}t}-\Delta_{EW}$, $\Delta m_{\tilde{t}t}-\Delta_{BG}$, and $\Delta_{BG}-\Delta_{EW}$ planes. The color coding is the same as Figure \ref{CMSSM_finetuning}. CMSSM with gravitino LSP has slightly better results in compare to those obtained for CMSSM with neutralino LSP; $\mu(\Lambda = {\rm weak})\sim 3-5$ TeV, $\Delta_{EW}\sim 2000-6000~(0.05\% - 0.017\%)$, and $\Delta_{BG}\sim 5000-15000~(0.02\% - 0.007\%)$.

%\vspace{-2.0cm}

\begin{table} 
\centering
\begin{tabular}{|c|ccccc|}
\hline
\hline
                 & Point 1 & Point 2 & Point 3 & Point 4 & Point 5\\

\hline
$m_{0}$         &6170  & 5753 & 5733 & 5244 & 5376\\
$M_{1/2} $      &1396 & 1198 & 1103 & 1298 & 1363\\
$\tan\beta$     &27.2 & 18.0 & 16.6 & 38.6 & 41.0\\
$A_0/m_{0}$     &2.80 & 2.71 & 2.65 & 2.92 & 2.95 \\
$m_t$           &{\color{red}173.3} & {\color{red}173.3} & {\color{red}173.3} & {\color{red}173.3} & {\color{red} 173.3} \\
\hline
$\mu$          & 5091 & 4678 & 4574 & 4455 & 4596\\
$\Delta a_{\mu}$ & $ { 0.13\times 10^{-11} } $ & $ { 0.94\times 10^{-11} } $ & $ { 0.87\times 10^{-11} } $ &  $ { 0.25\times 10^{-10} }$ & $ { 0.25\times 10^{-10} } $ 
\\ 
\hline
$m_h$            & {\color{red} 126.9}& {\color{red} 125.6} & {\color{red}125.6} & {\color{red} 125.9} & {\color{red} 126.1}\\
$m_H$           & 6980  & 7019 & 7012 & 5190 & 5125\\
$m_A$           & 6935 & 6973 & 6966 & 5156 & 5092\\
$m_{H^{\pm}}$   & 6981 & 7019 & 7012 & 5190 & 5126\\

\hline
$m_{\tilde{\chi}^0_{1,2}}$
                 & \textbf{615.8}, \textbf{1166}  & \textbf{ 526.8}, \textbf{1002}  & \textbf{ 483.8}, \textbf{922.1} & \textbf{567.2}, \textbf{1076} & \textbf{596.1}, \textbf{1130}\\

$m_{\tilde{\chi}^0_{3,4}}$
                 & 5127, 5128 & 4717, 4718  & 4619, 4620 & 4477, 4478 & 4613, 4614\\

$m_{\tilde{\chi}^{\pm}_{1,2}}$
                & \textbf{1180}, 5159 & \textbf{1014}, 4749 & \textbf{934.3}, 4652 & \textbf{1088}, 4502 & \textbf{1142}, 4637\\

$m_{\tilde{g}}$  & \textbf{3190} & \textbf{2784} & \textbf{2590} & \textbf{2960} & \textbf{3091}\\
\hline $m_{ \tilde{u}_{L,R}}$
                 & 6656, 6633  & 6138, 6123 & 6058, 6048 & 5750, 5724 & 5918, 5890\\
$m_{\tilde{t}_{1,2}}$
                 &{\color{red} 173.3}, 4328 & {\color{red} 157.9}, 4148 & {\color{red} 183.6}, 4111 & {\color{red} 169.4}, 3569 &  {\color{red} 172.9}, 3645\\
\hline $m_{ \tilde{d}_{L,R}}$
                 & 6656, 6633 & 6138, 6121  & 6058, 6046 & 5751, 5724 & 5919, 5886\\
$m_{\tilde{b}_{1,2}}$
                 & 4370, 6020  & 4187, 5797  & 4150, 5761 & 3599, 4921 & 3675, 5009\\
\hline
$m_{\tilde{\nu}_{e,\mu}}$
                 &6234 & 5803 & 5776 & 5309 & 5446\\
$m_{\tilde{\nu}_{\tau}}$
                 & 5597 &  5541 & 5758 & 4202 & 4156\\
\hline
$m_{ \tilde{e}_{L,R}}$
                & 6225, 6186 & 5794, 5765  & 5767, 5742 & 5302, 5262 & 5439, 5396\\
$m_{\tilde{\tau}_{1,2}}$
                & 4884, 5624 & 5249, 5550 & 5316, 5564 & 2705, 4238 & 2249, 3645\\
\hline
$\Delta_{EW}$ & 6463 & 5398 &  5163 &   4954 & 5258\\
$\Delta_{BG}$ & 15636 & 13389 & 13102 & 11576 & 12216 \\
\hline
\hline
\end{tabular}
\caption{Benchmark points for CMSSM with gravitino LSP. Masses are given in GeV unit. All points are chosen as to be consistent with the LHC bounds. Point 1 depicts a solution with exact degeneracy between stop and top quarks, Points 2 and 3 display solutions with the Higgs boson mass measured by ATLAS and CMS. Points 4 and 5 display solutions with different $\tan\beta$. } 
\label{benc_grav}
\end{table}

Finally we display five benchmark points that exemplify our results for CMSSM with gravitino LSP in Table \ref{benc_grav}. Masses are given in GeV unit. All points are chosen as to be consistent with the LHC bounds. Point 1 depicts a solution with exact degeneracy between stop and top quarks, Points 2 and 3 display solutions with the Higgs boson mass measured by ATLAS and CMS. Points 4 and 5 display solutions with different $\tan\beta$. The lesson from this section is that gravitino does not add much on NUHM1 in reducing the fine-tuning.

\section{CMSSM with $\mu< 0$ and Nonuniversal Gaugino Masses}

So far we have considered the stop-top degeneracy region in the CMSSM and NUHM1 for both neutralino LSP and gravitino LSP cases. The results show that it is possible to find such light stops whose pair production stays in error bar in calculation of the top quark pair production. While such regions are realized consistently with the experimental constraints, our results show that the models need to be highly fine-tuned. The best results for fine-tuning are obtained for the case with gravitino LSP ($\Delta_{EW}\sim 2000$), while it is worse for the cases with neutralino LSP in both CMSSM ($\Delta_{EW}\sim 9000$) and NUHM1 ($\Delta_{EW}\sim 8000$). If one excludes solutions requiring $\Delta_{EW} \lesssim 1000$, then stop-top degeneracy region in all the models considered in this paper disappears by the fine-tuning constraint.

\begin{figure}[h!]
\subfigure{\includegraphics[scale=1]{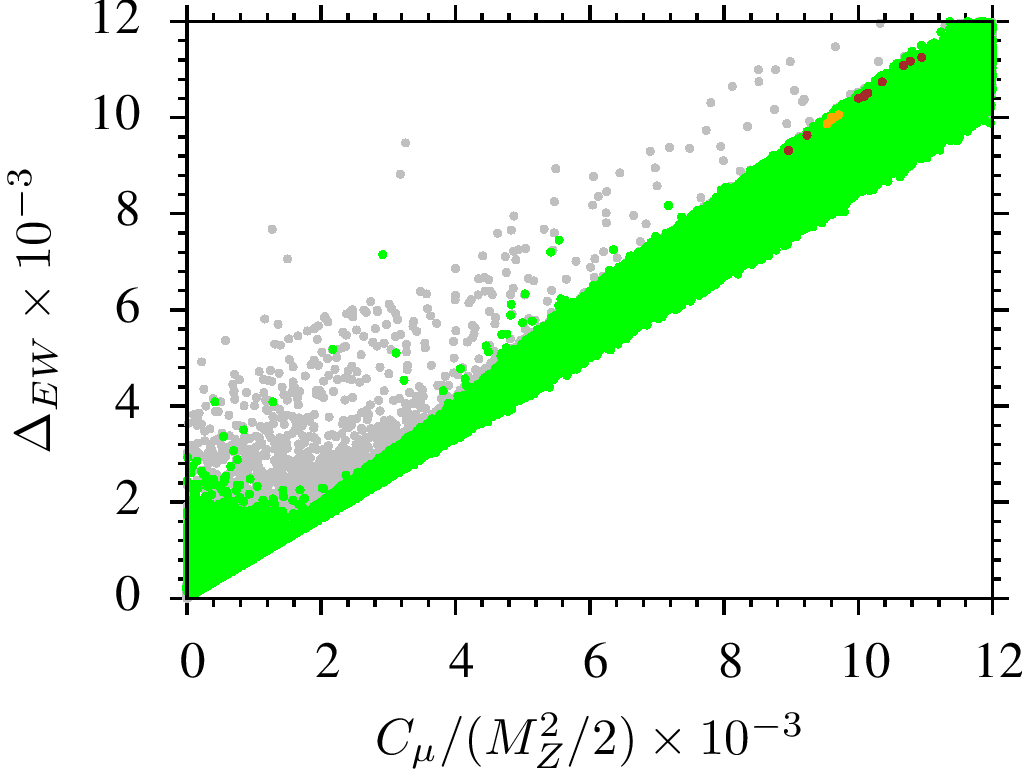}}
\subfigure{\includegraphics[scale=1]{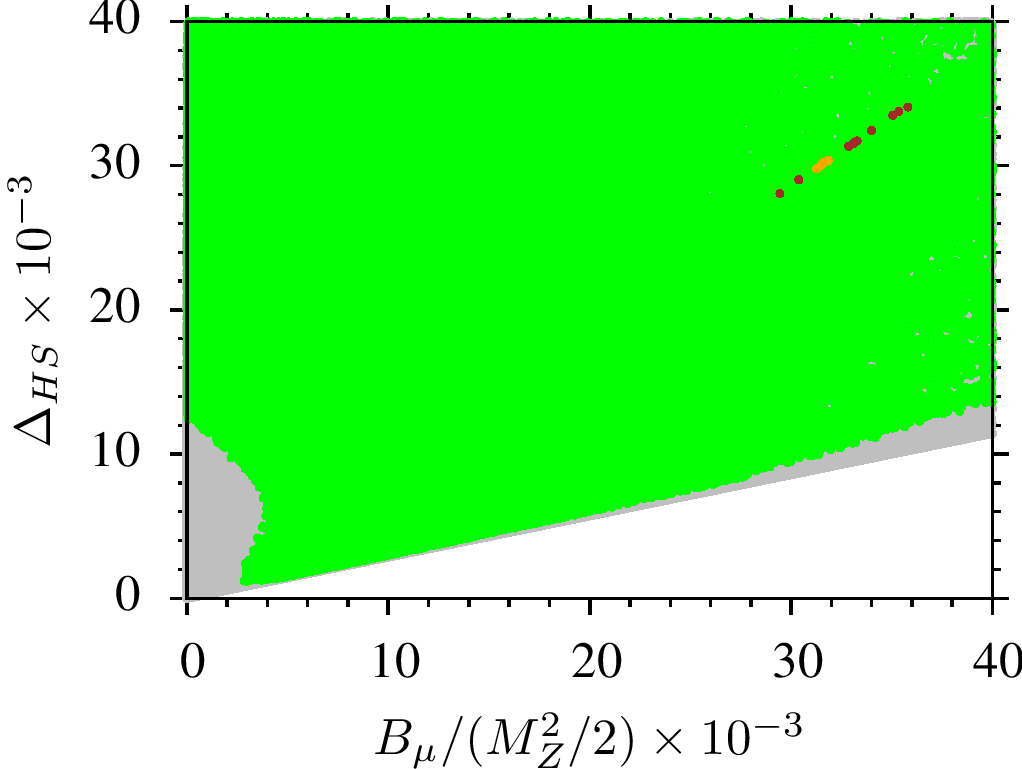}}
\caption{Plots in $\Delta_{EW}-C_{\mu}$ and $\Delta_{HS}-B_{\mu}$ planes. $C_{\mu}$ and $B_{\mu}$ are defined in Eqs.(4 and 6). All points are consistent with REWSB and neutralino LSP.Green points satisfy mass bounds and constraints from B-phsics, red points are a subset of green and they satisfy the $\Delta m_{\tilde{t}t} \leq 50$ GeV. Orange points is a subset of red and they yield relic density of neutralino LSP less than 1.}
\label{fineterms}
\end{figure}

In this section we discuss nonuniversality in gaugino masses. In general, fine-tuning indicates that there is a missing mechanism in the model under concern and its amount can be interpreted as effectiveness of the missing mechanism in the considered regions. There are exclusive studies on fine-tuning in supersymmetric models (for an incomplete list see \cite{Hall:2011aa}). The results from those studies show that fine-tuning constraint brings a lower bound on the stop mass around 500 GeV, even in  extensions of the MSSM. As discussed in Sec. III, IV and V, the stop-top degeneracy region requires a large mixing between stops. The connection between the fine-tuning and the large mixing can be established by considering the $\mu-$term. The mixing between the stops is proportional to $A_{t}-\mu \cot\beta$, and hence the $\mu-$term balances the contributions from large $A_{t}$ in order to adjust the stop mass such that it turns out to be nearly degenerate with the top quark. Besides, the $\mu-$term also determines the amount of fine-tuning as shown in plots in $\Delta_{EW}-C_{\mu}$ and $\Delta_{HS}-B_{\mu}$ planes of Figure \ref{fineterms}. All points in these plots are consistent with REWSB and neutralino LSP. Green points satisfy mass bounds and constraints from B-physics, red points are a subset of green and they satisfy the $\Delta m_{\tilde{t}t} \leq 50$ GeV. Orange points is a subset of red and they yield relic density of neutralino LSP less than 1. The linear correlations in Figure \ref{fineterms} means that $\Delta_{EW}$ is determined by $C_{\mu}$ and $\Delta_{BG}$ by $B_{\mu}$. 

\begin{figure}[h!]
\subfigure[$\mu > 0$]{\includegraphics[scale=1]{CMSSM_DeltaEW_DeltaHS.png}}
\subfigure[$\mu < 0$]{\includegraphics[scale=1]{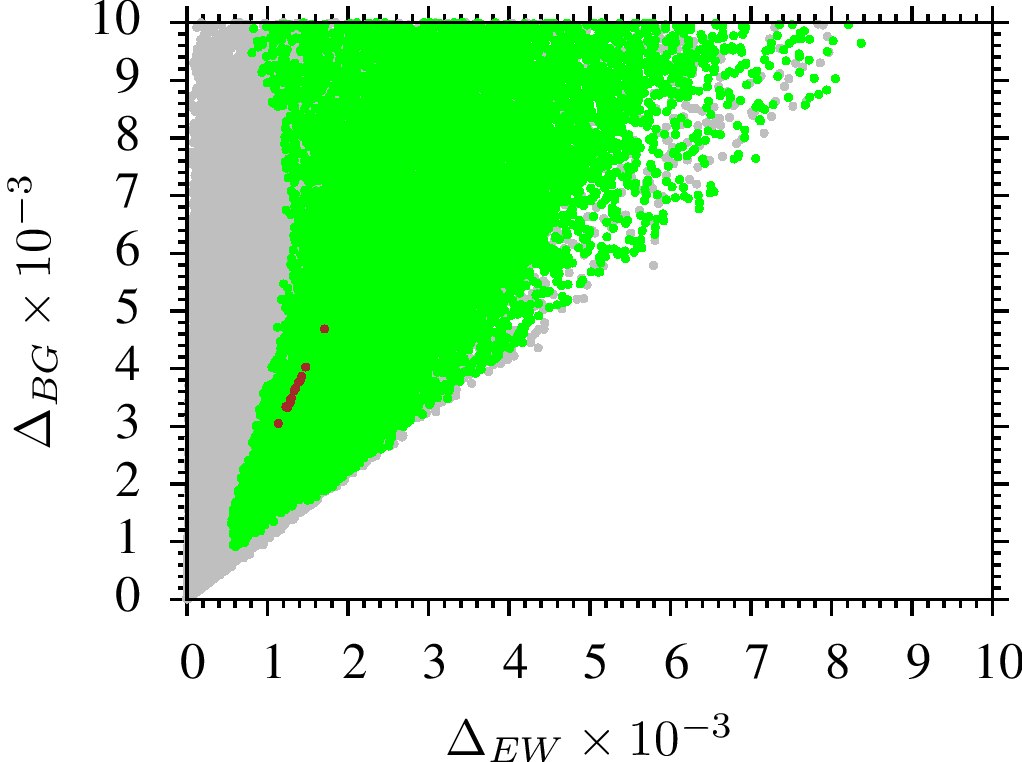}}
\caption{Plots in $\Delta_{BG}-\Delta_{HS}$ panels for CMSSM. The color coding is the same as Figure \ref{fineterms}. The left panel is for $\mu > 0$ and the right panel is for $\mu < 0$.} 
\label{munegmupos}
\end{figure}

The minimization of the Higgs potential allows both negative and positive signs for $\mu$, and hence, one can consider the case with negative $\mu$, which reverses the effect of $\mu$ in mixing of two stop quarks. The effect of the sign of $\mu$ on the fine-tuning is shown for CMSSM in $\Delta_{BG}-\Delta_{HS}$ panels of Figure \ref{munegmupos}. The negative $\mu$ is seen to reduce fine-tuning by an order of magnitude. 

Even though a significant improvement is realized in the case with $\mu < 0$ ($\Delta_{EW}\sim 2000$), it is still excluded by the constraint $\Delta_{EW} \lesssim 1000$. Can we  further lower fine-tuning? As we remember from non-universal Higgs masses, this can happen only if we increase the degrees of freedom. Namely, we must further deviate from CMSSM conditions. To this end, we explore the effects of nonunivesal gaugino masses. We study the CMSSM with $\mu< 0$ and nonuniversal gauginos and give the required fine-tuning for stop-top degeneracy region by scanning the model parameters in the following ranges:
\begin{eqnarray}
 0 \leq m_{0} \leq 20~{\rm TeV} \nonumber \\
  0 \leq M_{1} \leq 5~{\rm TeV} \nonumber \\
  0 \leq M_{2} \leq 5~{\rm TeV} \nonumber \\
  0 \leq M_{3} \leq 5~{\rm TeV}  \\
  -3 \leq A_{0}/m_{0} \leq 3 \nonumber \\
  2 \leq \tan\beta \leq 60 \nonumber \\
  \mu < 0 \nonumber
  \label{NUGM_param}
 \end{eqnarray} 
where sign of $\mu$ directly influences the stop left-right mixing, and hence, the mass splitting between the two mass eigenstates. 

\begin{figure}[ht!]
\subfigure{\includegraphics[scale=1]{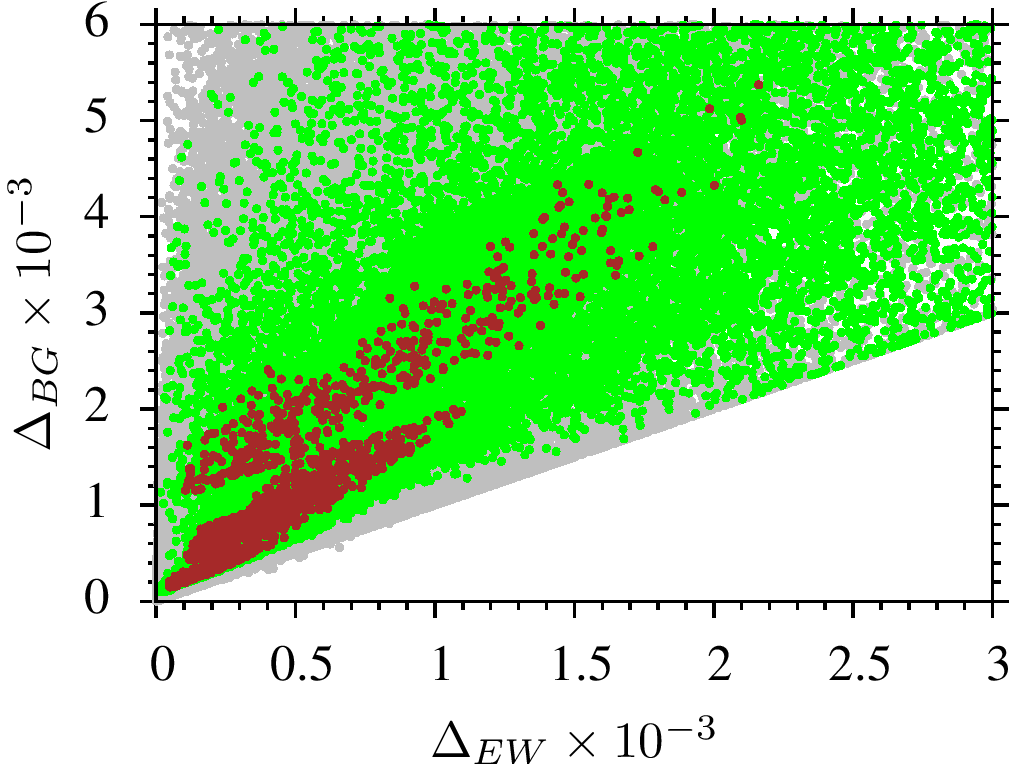}}
\subfigure{\includegraphics[scale=1]{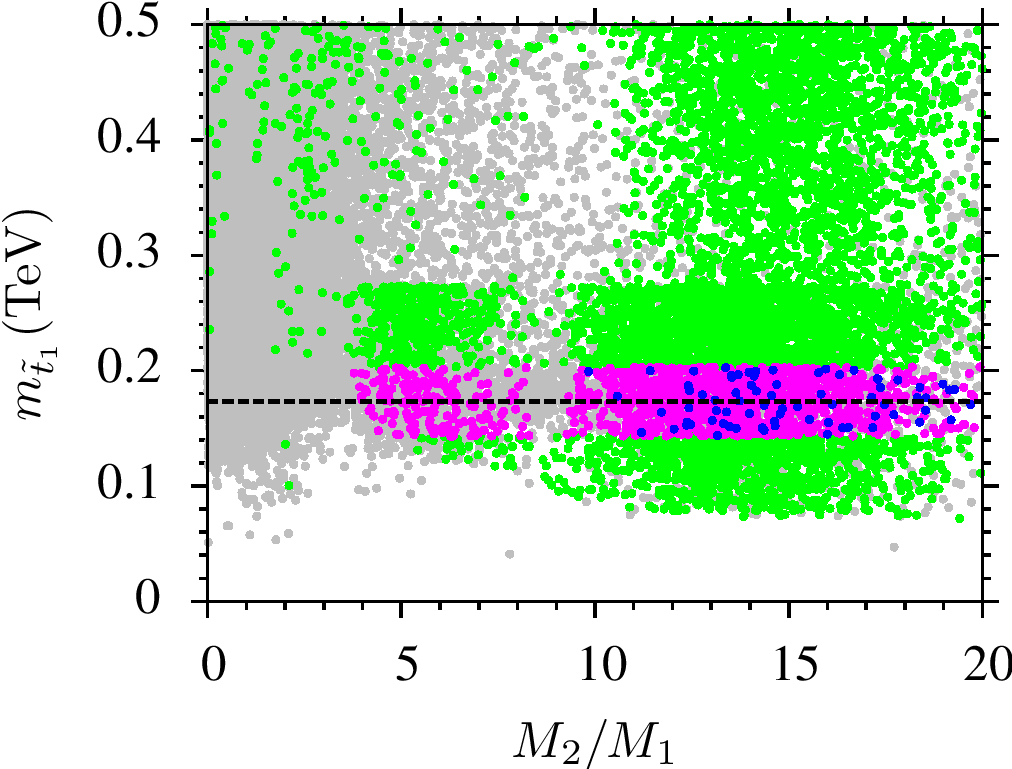}}
\subfigure{\includegraphics[scale=1]{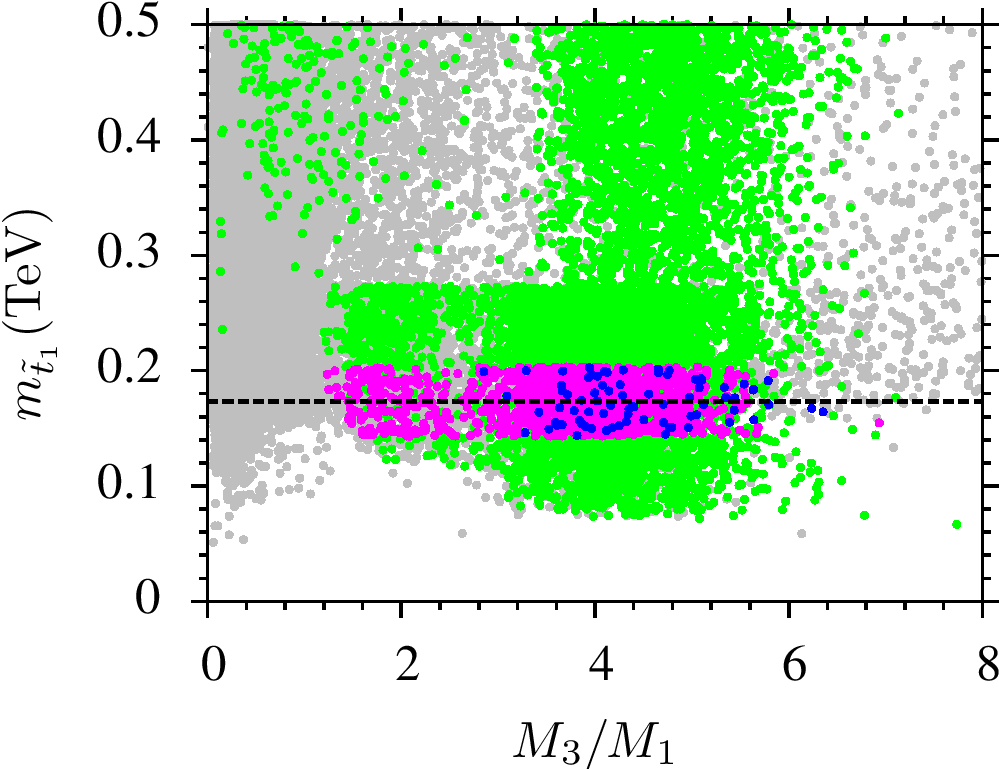}}
\subfigure{\hspace{1.4cm}\includegraphics[scale=1]{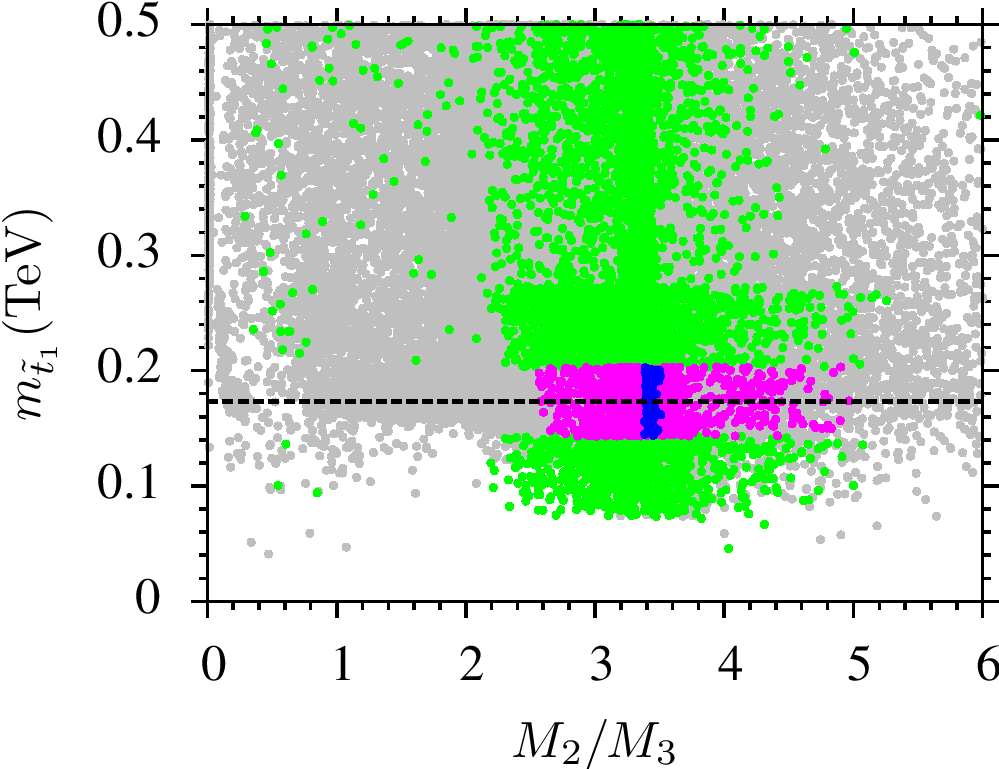}}
\caption{Plots in $\Delta_{BG} - \Delta_{EW}$, $m_{\tilde{t}_{1}} - M_{2}/M_{1}$, $m_{\tilde{t}_{1}}-M_{2}/M_{3}$, $m_{\tilde{t}_{1}}-M_{3}/M_{1}$ planes. All points are consistent with REWSB and neutralino LSP. Green points satisfy the mass bounds and constraints from B-physics. Red points are a subset of green and satisfy $\Delta m_{\tilde{t}t} \leq 30$ GeV. Magenta points form a subset of orange and represent the regions with $\Delta_{EW} \leq 500$. Similarly the blue points form a subset of magenta and they satisfy $\Delta_{EW} \leq 100$.}
\label{nugm_finetuning}
\end{figure}

Figure \ref{nugm_finetuning} represents plots in $\Delta_{BG} - \Delta_{EW}$, $m_{\tilde{t}_{1}} - M_{2}/M_{1}$, $m_{\tilde{t}_{1}}-M_{2}/M_{3}$, $m_{\tilde{t}_{1}}-M_{3}/M_{1}$ planes. All points are consistent with REWSB and neutralino LSP. Green points satisfy the mass bounds and constraints from B-physics. Orange points are a subset of green and satisfy $\Delta m_{\tilde{t}t} \leq 30$ GeV. Magenta points form a subset of orange and represent the regions with $\Delta_{EW} \leq 500$. Similarly the blue points form a subset of magenta and they satisfy $\Delta_{EW} \leq 100$. It can be seen that plenty of solutions for $\Delta_{EW} \leq 500$ can be realized in a model with nonuniversal gaugino masses at $M_{{\rm GUT}} $ and $\mu < 0$. From the plots in $m_{\tilde{t}_{1}} - M_{2}/M_{1}$ and $m_{\tilde{t}_{1}} - M_{3}/M_{1}$ planes we find the ratios for the gaugino masses as $M_{2}/M_{1} \gtrsim 5$ and $M_{3}/M_{1} \gtrsim 2$. As can be seen from the $m_{\tilde{t}_{1}} - M_{2}/M_{3}$, the ratio of $M_{2}$ to $M_{3}$ is found to be in a range $3 \lesssim M_{2}/M_{3} \lesssim 5$ for $\Delta_{EW} \leq 500$, while $M_{2}/M_{3} \approx 3.5$ if one applies the condition $\Delta_{EW} \leq 100$ strictly. It is clear that having $SU(2)$ gaugino so heavy compared to $SU(3)$ and $U(1)_Y$ gauginos blatantly violate the CMSSM conditions at high scale. In conclusion, compared to the CMSSM analyzed in Sec. III, its nonuniversal Higgs mass extension in Sec. IV, its supergravity structure in Sec. V, only the CMSSM $\mu<0$ extended by nonuniversal gaugino masses prove a viable reducing in fine-tuning. The fine-tuning cost of realizing stop-top degeneracy fall from ${\cal{O}}(10^{4})$ in the CMSSM down to ${\mathcal{O}}(100)$ with nonuniversal gaugino masses and $\mu < 0$.

\section{Conclusion}
\label{sec:conc}

In this work, we have searched for parameter regions which yield a light stop which is nearly degenerate with the top quark. Being the region not yet examined at the LHC, this narrow stripe is the place a
light stop can be hidden. Indeed, the LHC constraints for colored sparticles are severe and follow from non-observation of any of these sparticles. However, the cross section of stop pair production is so small that it is less than the experimental error bar in measurement of the top pair production. Namely,  light stops can hide in strong top quark background and it is difficult to distinguish them from the top quark signals (unless one performs precise spin measurements). 

We performed in this work a detailed search to determine if the CMSSM paramater space can accommodate a light stop nearly degenerate in mass with top quark. We explore degeneracy in a narrow band less than 50 GeV. In Sec. III, we consider the CMSSM parameter space with  $m_{0}\sim 9$ TeV, $M_{1/2}\sim 300$ GeV, $A_{0}/m_{0}\sim -2.2$, and $\tan\beta \sim 34$ while the squarks of the first two generations are kept  heavy ($\sim 9$ TeV) in agreement with the LHC constraints. The gluinos are found to be slightly above 1 TeV. On the other hand, the WMAP bound on the relic abundance of neutralino LSP excludes the regions with mass difference between NLSP stop and LSP neutralino up to 20$\%$ that make the degeneracy between stop and top worse. As we depicted in Sec. 3, stop-on-top scenario is realized by a fine-tuning ${\mathcal{O}}(10^{4})$ even when stop-top mass splitting is as large as 50 GeV. This fine-tuning is huge, and it tells us that there is something missing in modelling light scalar tops in the framework of the CMSSM.
 
We then start looking for extensions of the CMSSM in which fine-tuning can be lowered. We prefer to study cases where the CMSSM gauge group and particle spectrum are held. Namely, we do not study extended  models like next-to-MSSM. We start out exploration with non-universality in the mass parameters of the two Higgs doublets. The NUHM1 model is analyzed in Sec. IV and found to yield stop-on-top
with fine-tunings of ${\mathcal{O}}(10^{3})$. The CMSSM with gravitino LSP, as analyzed in Sec. V, requires similar order of fine-tuning. The CMSSM with $\mu< 0$ and nonuniveral gaugino masses however turns out to require much lower fine-tunings ${\mathcal{O}}(100)$. This price payed for this gain in fine-tuning is the nonuniversality in gaugino masses, where the $SU(2)$ gaugino is more massive than the $SU(3)$ gaugino. 
 
To sum up, the light stop band allowed by the LHC data is consistently realized in known models of supersymmetry at the expense of severe fine-tunings. The reason for fine-tuning is that stops must be light enough to facilitate a natural Higgs boson yet heavy enough to facilitate a heavy Higgs boson. Among conservative extensions of the CMSSM, only the one with $\mu< 0$ and nonuniversal gaugino masses gives least fine-tunings ${\mathcal{O}}(100)$. Further studies on CMSSM extensions can reveal more islands in the parameter space where stop-on-top scenario is naturally realized. 

\section*{Acknowledgment}
We would like to thank Qaisar Shafi and Liucheng Wang for fruitful discussions. We thank also Florian Staub and Ben O'Leary for useful comments concerning CCB minima. This work is supported in part by the TAEK CERN project CERN-A5.H2.P1.01-21 (DAD). This work used the Extreme Science and Engineering Discovery Environment (XSEDE), which is supported by the National Science Foundation Grant No. OCI-1053575, and part of computations have been carried out at ULAKBIM high performance center.

%\newpage

%\include{bibliography}

\end{document}